\documentclass[a4paper,journal]{IEEEtran}

\usepackage{etex}

\usepackage[font=footnotesize,caption=false]{subfig} 

\usepackage{relsize}

\usepackage{array}
\usepackage{booktabs,tabularx}
\usepackage{multirow}
\usepackage{units}
\usepackage{graphicx}

\usepackage[T1]{fontenc}
\usepackage{textcomp}
\usepackage[utf8]{inputenc}
\usepackage[final]{microtype}
\usepackage{icomma}
\usepackage{xspace}

\usepackage[tbtags]{amsmath}
\usepackage{amssymb,amsfonts,bm}
\usepackage{mathtools} 
\usepackage{dsfont}
\usepackage{mathrsfs}
\usepackage{accents}
\usepackage{empheq}
\usepackage{nccmath}
\usepackage{balance}

\usepackage{color}
\usepackage{calc}
\usepackage{tikz}
\usepackage{pgfplots,pgfplotstable}

\usepackage{setspace}

\usepackage{url}
\usepackage[capitalize]{cleveref}
\usepackage{refcount}

\usepackage[inline,shortlabels]{enumitem}
\usepackage{algorithm}
\usepackage{algpseudocode}
\makeatletter
\newcommand{\algmargin}{\the\ALG@thistlm}
\makeatother
\newlength{\whilewidth}
\algdef{SE}[parWHILE]{parWhile}{EndparWhile}[1]
  {\parbox[t]{\dimexpr\linewidth-\algmargin}{%
     \hangindent\whilewidth\strut\algorithmicwhile\ #1\ \algorithmicdo\strut}}{\algorithmicend\ \algorithmicwhile}%
\algnewcommand{\parState}[1]{\State%
  \parbox[t]{\dimexpr\linewidth-\algmargin}{\strut #1\strut}}

\usepackage{glossaries}
\usepackage{ifthen}
\usepackage{cite}
\usepackage{multibib}
\usepackage[shortcuts]{extdash}

\usepackage{comment}
\usepackage{todonotes}
\let\legacytodo\todo
\newcommand{\ruggedtodo}[2][]{\tikzexternaldisable\legacytodo[#1]{#2}\tikzexternalenable}
\renewcommand{\todo}[1]{\ruggedtodo[inline]{#1}}

\makeatletter
\def\todoref{\@ifnextchar[{\todoref@with}{\todoref@without}}
\def\todoref@without{\textbf{\color{red} [reference needed]}\xspace}
\def\todoref@with[#1]{\textbf{\color{red} [reference needed: #1]}\xspace}
\makeatother

\makeglossaries

\newacronym{dsl}{DSL}{digital subscriber line}
\newacronym{wsee}{WSEE}{weighted sum energy efficiency}
\newacronym{wpee}{WPEE}{weighted product energy efficiency}
\newacronym{wmee}{WMEE}{weighted minimum energy efficiency}
\newacronym{mmwave}{mmWave}{millimeter wave}
\newacronym{dfg}{DFG}{Deutsche Forschungsgemeinschaft}
\newacronym{haec}{HAEC}{Highly Adaptive Energy-Efficient Computing}
\newacronym{hpc}{HPC}{High Performance Computing}
\newacronym{mac}{MAC}{multiple-access channel}
\newacronym{bc}{BC}{broadcast channel}
\newacronym{siso}{SISO}{single-input single-output}
\newacronym{simo}{SIMO}{single-input multiple-output}
\newacronym{miso}{MISO}{multiple-input single-output}
\newacronym{mimo}{MIMO}{multiple-input multiple-output}
\newacronym{af}{AF}{amplify-and-forward}
\newacronym{df}{DF}{decode-and-forward}
\newacronym{cf}{CF}{compress-and-forward}
\newacronym{mwrc}{MWRC}{multi-way relay channel}
\newacronym{dmmwrc}{DM-MWRC}{discrete memoryless multi-way relay channel}
\newacronym{pde}{PDE}{partial data exchange}
\newacronym{fde}{FDE}{full data exchange}
\newacronym{iid}{i.i.d.\@}{independent and identically distributed}
\newacronym{di}{DI} {difference of increasing}
\newacronym{dc}{DC}{difference of convex}
\newacronym{awgn}{AWGN}{additive white Gaussian noise}
\newacronym{awg}{AWG}{additive white Gaussian}
\newacronym{sic}{SIC}{successive interference cancellation}
\newacronym{snr}{SNR}{signal-to-noise ratio}
\newacronym{sinr}{SINR}{signal to interference plus noise ratio}
\newacronym{inr}{INR}{interference to noise ratio}
\newacronym{zf}{ZF}{zero-forcing}
\newacronym{mrt}{MRT}{maximum ratio transmission}
\newacronym{mrc}{MRC}{maximum ratio combining}
\newacronym{mmse}{MMSE}{minimum mean square error}
\newacronym{sud}{SUD}{single user decoding}
\newacronym{dof}{DoF}{degrees of freedom}
\newacronym{gdof}{GDoF}{generalized degrees of freedom}
\newacronym{nnc}{NNC}{noisy network coding}
\newacronym{dmn}{DMN}{discrete memoryless network}
\newacronym{csi}{CSI}{channel state information}
\newacronym{pmf}{pmf}{probability mass function}
\newacronym{cdf}{CDF}{cumulative distribution function}
\newacronym{dmic}{DM-IC}{discrete memoryless interference channel}
\newacronym{ic}{IC}{interference channel}
\newacronym{ee}{EE}{energy efficiency}
\newacronym{gee}{GEE}{global energy efficiency}
\newacronym{ian}{IAN}{treating interference as noise}
\newacronym{snd}{SND}{simultaneous non-unique decoding}
\newacronym{sd}{SD}{simultaneous decoding}
\newacronym{hk}{HK}{Han-Kobayashi}
\newacronym{rs}{RS}{rate splitting}
\newacronym{rf}{RF}{radio frequency}
\newacronym{pa}{PA}{power amplifier}
\newacronym{lna}{LNA}{low noise amplifier}
\newacronym{lo}{LO}{local oscillator}
\newacronym{adc}{ADC}{analog-to-digital converter}
\newacronym{dac}{DAC}{digital-to-analog converter}
\newacronym{dsp}{DSP}{digital signal processing}
\newacronym{brd}{BRD}{best response dynamics}
\newacronym{br}{BR}{best response}
\newacronym{ne}{NE}{Nash equilibrium}
\newacronym{lhs}{LHS}{left-hand side}
\newacronym{rhs}{RHS}{right-hand side}
\newacronym{ran}{RAN}{radio access network}
\newacronym{qos}{QoS}{Quality of Service}
\newacronym{ngmn}{NGMN}{Next Generation Mobile Networks}
\newacronym{cap}{CAP}{Capacity Adaptation}
\newacronym{bwa}{BW}{Bandwidth Adaptation}
\newacronym{prb}{PRB}{physical resource block}
\newacronym{se}{SE}{spectral efficiency}
\newacronym{tp}{TP}{throughput}
\newacronym{bs}{BS}{base station}
\newacronym{ue}{UE}{user equipment}
\newacronym{mop}{MOP}{multi-objective optimization problem}
\newacronym{gda}{GDA}{generalized Dinkelbach's algorithm}
\newacronym{midcp}{MIDCP}{mixed integer disciplined convex programming}
\newacronym{lp}{LP}{linear program}
\newacronym{brb}{BRB}{branch reduce and bound}
\newacronym{bb}{BB}{branch-and-bound}
\newacronym{sit}{SIT}{successive incumbent transcending}
\newacronym{oma}{OMA}{orthogonal multiple access}
\newacronym{noma}{NOMA}{non-orthogonal multiple access}
\newacronym{wlog}{w.l.o.g.\@}{without loss of generality}
\newacronym{lsc}{l.s.c.\@}{lower semi-continuous}
\newacronym{usc}{u.s.c.\@}{upper semi-continuous}
\newacronym{kkt}{KKT}{Karush-Kuhn-Tucker}
\newacronym{ann}{ANN}{artificial neural network}
\newacronym{sca}{SCA}{successive convex approximation}
\newacronym{cbv}{CBV}{current best value}
\newacronym{cbs}{CBS}{current best solution}

\glsenableentrycount
\makeglossaries

\usetikzlibrary{positioning}
\usetikzlibrary{calc}
\usetikzlibrary{fit}
\usetikzlibrary{intersections}
\usetikzlibrary{decorations.pathreplacing}
\usetikzlibrary{pgfplots.fillbetween}
\usetikzlibrary{bending}
\usetikzlibrary{arrows.meta}
\usetikzlibrary{decorations.markings}

\usetikzlibrary{external}

\pgfplotsset{compat=newest}

\pgfkeys{/pgfplots/.cd,
	hk/.style={blue,mark=*},
	snd/.style={red,mark=square*},
	ian/.style={brown!60!black,mark=triangle*},
}

\makeatletter
\newcommand\transformxdimension[1]{
    \pgfmathparse{((#1/\pgfplots@x@veclength)+\pgfplots@data@scale@trafo@SHIFT@x)/10^\pgfplots@data@scale@trafo@EXPONENT@x}
}
\newcommand\transformydimension[1]{
    \pgfmathparse{((#1/\pgfplots@y@veclength)+\pgfplots@data@scale@trafo@SHIFT@y)/10^\pgfplots@data@scale@trafo@EXPONENT@y}
}
\makeatother

\crefname{equation}{}{}
\crefrangeformat{equation}{(#3#1#4)--(#5#2#6)}
\crefformat{footnote}{#2\footnotemark[#1]#3}

\newcommand{\abs}[1]{\ensuremath{\left\lvert #1 \right\rvert}}
\newcommand{\norm}[1]{\ensuremath{\left\lVert #1 \right\rVert}}

\DeclareMathOperator*{\argmax}{arg\,max}

\let\vec\bm

\allowdisplaybreaks[3]

\DeclareFontFamily{U}{mathx}{\hyphenchar\font45}
\DeclareFontShape{U}{mathx}{m}{n}{
      <5> <6> <7> <8> <9> <10>
      <10.95> <12> <14.4> <17.28> <20.74> <24.88>
      mathx10
      }{}
\DeclareSymbolFont{mathx}{U}{mathx}{m}{n}
\DeclareMathSymbol{\bigtimes}{1}{mathx}{"91}

\newtheorem{proposition}{Proposition}

\newtheorem{remark}{Remark}

\newcounter{optimizationproblem}
\newenvironment{optprob}{\begin{equation}\left\{\begin{aligned}}{\end{aligned}\right.\refstepcounter{optimizationproblem}\tag{P\theoptimizationproblem}\end{equation}\ignorespacesafterend}
\newenvironment{optprob*}{\begin{equation*}\left\{\begin{aligned}}{\end{aligned}\right.\end{equation*}\ignorespacesafterend}

\makeatletter
\newcommand{\subalign}[1]{%
  \vcenter{%
    \Let@ \restore@math@cr \default@tag
    \baselineskip\fontdimen10 \scriptfont\tw@
    \advance\baselineskip\fontdimen12 \scriptfont\tw@
    \lineskip\thr@@\fontdimen8 \scriptfont\thr@@
    \lineskiplimit\lineskip
    \ialign{\hfil$\m@th\scriptstyle##$&$\m@th\scriptstyle{}##$\crcr
      #1\crcr
    }%
  }
}
\makeatother

\pgfplotscreateplotcyclelist{default}{%
	blue,mark=*\\%
	red,mark=star\\%
	teal,mark=square*\\%
	brown!60!black,mark=otimes*\\%
}

\pgfplotscreateplotcyclelist{long}{%
	red,solid,\\%
	blue,dashed,\\%
	black,dashdotted,\\%
	teal,densely dotted,\\%
	brown!60!black,dashdotdotted\\%
	cyan,densely dashdotted\\%
}

\makeatletter
\newif\ifhbonecolumn
\@ifclasswith{IEEEtran}{onecolumn}{\hbonecolumntrue}{\hbonecolumnfalse}
\makeatother

\begin{document}
\bstctlcite{IEEEexample:BSTcontrol}
\title{A Globally Optimal Energy-Efficient Power Control Framework and its Efficient Implementation in Wireless Interference Networks}

\author{Bho~Matthiesen,~\IEEEmembership{Student Member,~IEEE,} Alessio Zappone,~\IEEEmembership{Senior Member,~IEEE,} Karl-L.~Besser,~ \IEEEmembership{Student  Member,~IEEE,} Eduard~A.~Jorswieck,~\IEEEmembership{Senior Member,~IEEE,} Merouane Debbah,~\IEEEmembership{Fellow,~IEEE}  
		\thanks{
			Parts of this paper were presented at IEEE WCNC 2018 \cite{wcnc18} and IEEE SPAWC 2019 \cite{spawc19a}.
		}%
		\thanks{
			B.~Matthiesen is with the Chair for Communications Theory, Technische Universität Dresden, Dresden, Germany (e-mail: bho.matthiesen@tu-dresden.de).
			A.~Zappone is with DIEI, University of Cassino and Southern Lazio, Cassino, FR, Italy (e-mail: alessio.zappone@unicas.it).
			K.-L.~Besser and E.~A.~Jorswieck are with the Department of Information Theory and Communication Systems, TU Braunschweig, Germany (e-mail: k.besser@tu-bs.de, e.jorswieck@tu-bs.de).
			M. Debbah is with Université Paris-Saclay, CNRS, CentraleSupélec,  Laboratoire des signaux et systèmes, 91190, Gif-sur-Yvette, France (e-mail: merouane.debbah@l2s.centralesupelec.fr) and with the Mathematical and Algorithmic Sciences Laboratory, France Research Center, Huawei Technologies, Paris, France.
		}%
	}

\maketitle

\begin{abstract}
This work develops a novel power control framework for energy-efficient power control in wireless networks. The proposed method is a new \acrlong{bb} procedure based on problem-specific bounds for energy-efficiency maximization that allow for faster convergence. This enables to find the global solution for all of the most common energy-efficient power control problems with a complexity that, although still exponential in the number of variables, is much lower than other available global optimization frameworks. Moreover, the reduced complexity of the proposed framework allows its practical implementation through the use of deep neural networks. Specifically, thanks to its reduced complexity, the proposed method can be used to train an artificial neural network to predict the optimal resource allocation. This is in contrast with other power control methods based on deep learning, which train the neural network based on suboptimal power allocations due to the large complexity that generating large training sets of optimal power allocations would have with available global optimization methods. As a benchmark, we also develop a novel first-order optimal power allocation algorithm. Numerical results show that a neural network can be trained to predict the optimal power allocation policy.
\end{abstract}
\glsresetall


\section{Introduction}
Energy management is known to be one of the crucial issues for the sustainability of future wireless communication networks, whose bit-per-Joule \gls{ee} is required to increase by a factor 2000 compared to present networks \cite{5GNGMN}. To this end, several energy management techniques have been proposed, such as energy-efficient network deployment, the use of renewable energy sources, as well as the development of resource allocation techniques aimed at \gls{ee} maximization \cite{GEJSAC16}. This work focuses on the last of these energy management approaches, and in particular on  the issue of energy-efficient power control. 

Due to the fractional nature of energy-efficient performance metrics, traditional convex optimization theory can not directly handle energy-efficient power control problems. Instead, the mathematical frameworks of generalized concavity theory and fractional programming provide a suitable set of optimization methods. However, these optimization tools come with a limited complexity only when the fractional function to maximize fulfills certain mathematical assumptions, such as the concavity of the numerator \cite{ZapNow15}. Unfortunately, this requirement is not fulfilled whenever an interference-limited network needs to be optimized, and indeed in these cases \gls{ee} maximization problems are typically NP-hard \cite{Luo2008}. Thus, several suboptimal methods have been proposed for energy-efficient resource allocation in wireless interference networks. The simplest approach is to resort to interference cancelation techniques or to orthogonal transmission schemes, thus falling back into the noise-limited regime \cite{ng_limitedbackhaul2012,xu2014het,MIMOBroadcastEE}. However, this either leads to a poor resource efficiency, or to noise enhancement effects and/or non-linear receive schemes. A more recent approach is instead of trying to develop energy-efficient power control algorithms that, although not provably optimal, enjoy limited (typically polynomial) complexity. In \cite{he2013coordinated}, the maximization of the system \gls{gee} is pursued by merging fractional programming with alternating optimization, decomposing the problem into a series of simpler sub-problems. A similar approach is used in \cite{Du14}, where the minimum of the users' \glspl{ee} is maximized, and in \cite{he2014coordinated}, where the sum of the individual  users' \glspl{ee} is considered. In \cite{ZapTSP15}, fractional programming is merged with sequential optimization theory to develop power control algorithms for the maximization of the system \gls{gee} or minimum of the users' \glspl{ee}. Unlike previous contributions, the method proposed there guarantees first-order optimality, and has been numerically shown to achieve global optimality in small network setups in \cite{ZapTSP17}. Results for networks with a larger number of users are not available at present, mainly due to the fact that, in order to compute the global maximum of the \gls{gee} in networks with more than a handful of users, available global optimization frameworks require a computational complexity that is impractical even for offline simulation. This issue is even more severe when other energy-efficient metrics are considered, such as the product or the sum of the users' \glspl{ee}. In particular, the maximization of the sum of the users'  \glspl{ee} is acknowledged as the hardest energy-efficient power control problem \cite{ZapNow15}, and it has been numerically shown in \cite{wcnc18} that, unlike what was shown for the \gls{gee} in \cite{ZapTSP17}, state-of-the-art first-order optimal methods with polynomial complexity have a gap to the global solution even in networks with a few users. Moreover, it is known that \gls{wsee} maximization is an NP-complete optimization problem even if each ratio has concave numerator and linear denominator \cite{Freund2001}, due to the fact that addition is not guaranteed to preserve properties like pseudo-concavity or quasi-concavity.

Thus, the analysis of the state-of-the-art shows a significant gap regarding the availability of  optimization frameworks that allow the computation of the global maximum of energy-efficient performance metrics in wireless interference networks with a complexity that is at least affordable for offline simulation. Besides having its own theoretical value, developing a framework for efficient offline global optimization is of interest also because it provides a practical way to benchmark the performance of any sub-optimal, low-complexity optimization routine. Finally, having efficient global optimization algorithms represents a key requirement also for the use of deep learning in wireless communications, a topic that has been gaining momentum recently \cite{OShea17,Chen2017}. Indeed, an efficient global optimization algorithm provides a feasible way to generate large training sets, which is a critical requirement in order for  \glspl{ann} to perform well.

Specifically, the main contributions of the work are as follows:
\begin{itemize}
\item We develop a novel and improved \cgls{bb}-based algorithm to globally solve the most common energy-efficient power control problems with a complexity that is much lower than available global optimization approaches for these kind of problems. This is achieved thanks to the development of improved bounding techniques that significantly accelerate the convergence of the algorithm to the global solution. This makes it possible to globally solve NP-hard optimization problems in a time that is fully affordable for offline simulations.
\item In addition to the newly proposed \cgls{bb}-based algorithm, we also propose a novel power control algorithm that is guaranteed to obtain a first-order optimal solution of the \gls{wsee} maximization problem with polynomial complexity. Although not enjoying global optimality, this approach provides a practical benchmark to assess the performance of other optimization methods.
\item Finally, we also develop a feedforward \gls{ann}-based energy-efficient power control method that works in tandem with the proposed \cgls{bb}-based algorithm and leverages the universal approximation property of \glspl{ann}. Specifically, the lower complexity of the proposed \cgls{bb}-based algorithm enables the offline generation a large training set containing optimal power allocations for many different channel realizations, which is then used to train an \gls{ann} to learn the optimal map between the network channel realization and the corresponding optimal power control policy. Afterwards, the trained \gls{ann} is able to provide a power control policy for new realizations of the network channels not contained in the training set with an extremely limited computational complexity that essentially amounts to performing one forward propagation of the \gls{ann}. Thus, the trained \gls{ann} can be applied as an effective and low-complexity online power allocation method. It is to be stressed how this approach differs from previous related works, which train the \gls{ann} based on suboptimal power allocation routines \cite{Sun2017,Liang2018}, and are thus intrinsically limited by the sub-optimality of the data in the training set. Instead, the reduced complexity of our proposed \cgls{bb}-based algorithm makes it practical to generate offline large training sets with optimal power allocations. 
\item Extensive numerical results are provided to assess the performance of the proposed approaches. Interestingly, it is found that both the proposed first-order approach and the \gls{ann}-based approach achieve near-optimal performance but at very different computational cost. Moreover, our numerical analysis shows that the proposed \gls{ann}-based method is even robust to mismatches in the channel statistics between the training data and the actual scenario in which the \gls{ann} is tested. 
\end{itemize}
The rest of the work is organized as follows. \Cref{Sec:SystemModel} describes the system model and formulates the power control problem. \Cref{Sec:BB} introduces the proposed globally optimal algorithm. \Cref{Sec:SEQ} introduces the proposed first-order optimal power control algorithm, while \cref{Sec:ANN} develops the \gls{ann}-based power control method that employs an \gls{ann} trained on the globally optimal algorithm from \cref{Sec:BB}. \Cref{sec:numeval} illustrates the numerical performance assessment of the proposed algorithms, while concluding remarks are provided in \cref{Sec:Conclusions}. 

\textbf{Notation:} Vectors $\vec a$ and matrices $\vec A$ are typeset in bold face, sets and maps in calligraphic letters $\cal A$.
The sets of real and complex numbers are denoted by $\mathds R$ and $\mathds C$, respectively.
We define the sets $\{ a_i \}_{i = 1}^n = \{ a_1, a_2, \dots, a_n \}$ and vectors $( a_i )_{i = 1}^n = ( a_1, a_2, \dots, a_n )$ where we might omit the index bounds if clear from the context. $[\vec a, \vec b] = \{ \vec x \,|\, a_i \le x_i \le b_i, \text{ for all } i = 1, \dots, n\}$, while $\vec a \le \vec b$ means $a_i \le b_i$ for all $i = 1, \dots, n$, $\vec 0$ denotes a zero vector of appropriate dimension, and scalar functions applied to vectors are to be applied element-wise, e.g., $\log\vec a = ( \log a_i )_{i=1}^n$.
The operators $\norm{\cdot}$, $\abs{\cdot}$, $(\cdot)^T$, and $(\cdot)^H$ denote the $L^2$-norm, absolute value, transpose, and conjugate transpose, respectively.
Logarithms are, unless noted otherwise, to the base 2, and $\ln$ denotes the natural logarithm.
Definitions and approximations are marked by $\coloneqq$ and $\approx$, respectively.

\section{System Model \& Problem Statement}\label{Sec:SystemModel}
Consider the uplink of a multi-cell interference network with $L$ single antenna \cglspl{ue} served by $M$ \cgls{bs}, equipped with $n_R$ antennas each. Let $a(i)$ be the \gls{bs} serving user $i$. Then, the received signal at \cgls{bs} $a(i)$ is
\begin{equation}\label{Eq:RecSignal}
	\vec y_{a(i)} = \sum_{j=1}^{L} \vec h_{a(i),j} x_j + \vec z_{a(i)}
\end{equation}
wherein $\vec h_{a(i),j}\in\mathds C^{n_R}$ is the channel from \cgls{ue} $j$ to \cgls{bs} $a(i)$, $x_j\in\mathds C$ the symbol transmitted by \cgls{ue} $j$, and $\vec z_{a(i)}$ zero-mean circularly symmetrical complex Gaussian noise with power $\sigma_i^2$. Each \cgls{ue} is subject to an average transmit power constraint, i.e., $p_j \le P_j$ where $p_j$ is the average power of $x_j$. Upon matched-filter reception, 
 and under the assumption of Gaussian codebooks, the achievable rate from \cgls{ue} $i$ to its intended  \cgls{bs} is
\begin{equation}\label{Eq:Rate}
	R_i = B \log\left( 1 + \frac{\alpha_i p_i}{1 + \sum_{j\neq i} \beta_{i,j} p_j} \right)
\end{equation}
with $B$ the communication bandwidth, $\alpha_i = \frac{\norm{\vec h_{a(i),i}}^2}{\sigma_i^2}$, and $\beta_{i,j} = \frac{\abs{\vec h_{a(i),i}^H \vec h_{a(i),j}}^2}{\sigma_i^2 \norm{\vec h_{a(i),i}}^2}$.

In this context, the \cgls{ee} of the link between \cgls{ue} $i$ and its intended \cgls{bs} is defined as the benefit-cost ratio in terms of the link's achievable rate and power consumption necessary to operate the link, i.e.,
\begin{equation} \label{eq:eei}
\text{EE}_i = \frac{\displaystyle B \log\left( 1 + \frac{\alpha_i p_i}{1 + \sum_{j\neq i} \beta_{i,j} p_j} \right)}{\mu_i p_i + P_{c,i}}
\end{equation}
with $\mu_i$ the inefficiency of \cgls{ue} $i$'s power amplifier and $P_{c,i}$ the total static power consumption of \cgls{ue} $i$ and its associated \cgls{bs}. 

\subsection{Problem Formulation and Motivation}\label{Sec:Motivation}
The goal of this work is to develop an efficient and effective algorithm to solve energy-efficient power control problems. The four fundamental energy-efficient metrics that have received most research attention in the open literature are the \gls{gee}, the \gls{wsee}, the \gls{wpee}, and the \gls{wmee},  defined as 
\begin{align}\label{Eq:GEE}
	\text{GEE}&=\frac{\sum_{i=1}^{L}\log\left(1 + \frac{\alpha_i p_i}{1 + \sum_{j\neq i} \beta_{i,j} p_j} \right)}{\sum_{i=1}^{L}\left( \mu_i p_i + P_{c,i} \right)}\\
\label{Eq:WSEE}
\text{WSEE}&=\sum_{i=1}^L  w_{i}\frac{\log\left( 1 + \frac{\alpha_i p_i}{1 + \sum_{j\neq i} \beta_{i,j} p_j} \right)}{\mu_i p_i + P_{c,i}}\\
\label{Eq:WPEE}
\text{WPEE}&=\prod_{i=1}^L  \left(\frac{\log\left( 1 + \frac{\alpha_i p_i}{1 + \sum_{j\neq i} \beta_{i,j} p_j} \right)}{\mu_i p_i + P_{c,i}}\right)^{\!w_{i}}\\
\label{Eq:WMEE}
\text{WMEE}&=\min_{1\leq i\leq L} w_{i}\frac{\log\left( 1 + \frac{\alpha_i p_i}{1 + \sum_{j\neq i} \beta_{i,j} p_j} \right)}{\mu_i p_i + P_{c,i}}\;.
\end{align}
In order to consider a specific case-study, the optimization algorithms to be developed in the following will be stated primarily with reference to the problem of \gls{wsee} maximization. Nevertheless, it must be stressed that both the global optimization framework and the \gls{ann}-based method to be developed  are general enough to apply to all four energy-efficient metrics above. This will be explicitly shown in \cref{Sec:BB} with reference to the global optimization method, and in \cref{Sec:ANN} with reference to the \gls{ann}-based method. 

Regarding the motivation for choosing the \gls{wsee} as main case-study, several arguments can be made. First of all, as already mentioned, the \gls{wsee} is the hardest to maximize among energy-efficient metrics. Thus, showing that our optimization framework performs well in this case represents a strong motivation for its use to tackle other energy-efficient power control problems, too. Among the reasons that make Problem \cref{opt} so challenging to handle by traditional optimization techniques, the following observations hold:
\begin{itemize}
\item The objective of \cref{opt} is a sum of fractions, a functional form that is NP-complete in general \cite{Freund2001}, and which indeed can not be tackled with polynomial complexity by any available fractional programming technique. 
\item Each summand of the objective of \cref{opt} is a fraction with a non-concave numerator, which would make \cref{opt} NP-hard even if only the weighted sum-rate were to be maximized (i.e., if $\mu_{i}=0$ for all $i=1, \ldots, L$) \cite{Luo2008}. 
\end{itemize}
Another reason to motivate the consideration of the \cgls{wsee} lies in its different operational meaning compared to the more widely-considered \gls{gee} metric. While the latter is meant to optimize the \cgls{ee} of the network as a whole, the former enables to balance the \cgls{ee} levels among the \cglspl{ue}, thanks to the use of the weights $w_i\ge 0$, prioritizing the \cgls{ee} of some links over others. For example, some terminals might not have a reliable energy source, e.g., due to being powered by a renewable energy source. In this case, it can be useful to prioritize the individual \cgls{ee} of these users over that of the other users. 

Further elaborating on the prioritization of the users' individual \cglspl{ee}, it is important to mention that Problem~\eqref{opt} can be cast into the framework of multi-objective optimization. Formally speaking, let us consider the problem of jointly maximizing all of the users' individual \cglspl{ee}, namely
\begin{equation}\label{Eq:MOP}
\max_{\vec 0\le \vec p \le \vec P} [ \text{EE}_{1}(\vec p), \text{EE}_{2}(\vec p), \ldots, \text{EE}_{L}(\vec p)]
\end{equation}
with $\vec p=[p_{1}, \ldots, p_{L}]$ and $\vec P=[P_{1}, \ldots, P_{L}]$. A moment's thought shows that the individual \cglspl{ee} are conflicting objectives, since, for all $i$, \cref{eq:eei} is strictly decreasing in $p_j$ for all $j \neq i$, i.e., $p_j = 0$ for all $j\neq i$ maximizes \cref{eq:eei}, but the optimal $p_i$ is strictly positive. Thus, no power allocation vector exists that simultaneously maximizes all individual \cglspl{ee}. In this context, in order to define a suitable solution concept for \eqref{Eq:MOP}, the notion of energy-efficient Pareto region is defined as the set of all \gls{ee} vectors $[\text{EE}_{1}(\vec p), \text{EE}_{2}(\vec p), \ldots, \text{EE}_{L}(\vec p)]$ which can be attained by a feasible power allocation vector $\vec p$. The outer boundary of the Pareto region is called Pareto boundary, and provides all feasible operating points at which it is not possible to improve the \cgls{ee} of one user, without decreasing the \cgls{ee} of another user \cite{Zadeh1963,Miettinen1999}. For this reason, the  points on the Pareto-boundary are called Pareto-optimal and are commonly understood as the solutions of the multi-objective optimization problem. 

Having said this, a popular method of determining Pareto-optimal solutions is the so-called scalarization approach, which consists of maximizing a weighted sum of the objectives. Therefore, Problem \cref{opt} is  the scalarized version of the multi-objective Problem~\eqref{Eq:MOP}, and thus solving \cref{opt} yields a Pareto-optimal\footnote{It can be proved that solving a scalarized problem for varying combinations of the weights enables to obtain the complete convex hull of the Pareto-boundary.} solution of \eqref{Eq:MOP}.

A third motivation to focus on the \cgls{wsee} is that it is a direct generalization of the system weighted sum-rate 
\begin{equation}\label{Eq:WSR}
\text{WSR}=\sum_{i=1}^L  w_{i}\log\left( 1 + \frac{\alpha_i p_i}{1 + \sum_{j\neq i} \beta_{i,j} p_j} \right)\;,
\end{equation}
obtained from the \gls{wsee} by setting $\mu_{i}=0$ and $P_{c,i}=1$ for all $i=1, \ldots, L$. In addition, we emphasize that the global optimization approach and the \gls{ann}-based approach to be developed in  \cref{Sec:BB,Sec:ANN}, respectively, are not restricted to the \gls{wsee}, and will be shown to encompass all four major energy-efficient metrics defined above.

Finally, based on all the above considerations, the main optimization problem to be considered in the following sections is the maximization of the \cgls{wsee} subject to maximum power constraints, which is formulated as 
\begin{optprob}
	& \underset{\vec p}{\text{max}}
	&& \sum_{i=1}^L  w_{i}\frac{\log\left( 1 + \frac{\alpha_i p_i}{1 + \sum_{j\neq i} \beta_{i,j} p_j} \right)}{\mu_i p_i + P_{c,i}} \\
	& \text{s.\,t.}
	&& 0 \le p_i \le P_i,\quad\text{for all } i = 1, 2, \ldots, L,
	\label{opt}
\end{optprob}
The coming section develops a novel \gls{bb}-based method to obtain the global solution of \eqref{opt}, with a lower complexity than other global optimization frameworks. 

\section{Globally Optimal Power Control}\label{Sec:BB}
Problem \cref{opt} has, in general, multiple locally optimal solutions and is known to be NP-complete \cite{Freund2001}. Thus, traditional optimization approaches like gradient descent or interior-point methods are not able to solve \cref{opt} globally. Instead, in this section we develop a novel \cgls{bb} procedure to solve \cref{opt} with guaranteed global optimality, while at the same time lowering the complexity with respect to \cgls{bb} methods using general-purpose bounds, e.g., monotonic optimization \cite{Tuy2005} or mixed monotonic programming \cite{mmp}.

The approach successively partitions the set $[\vec 0, \vec P] = [0,P_{i}]^{L}$ into $L$-dimensional hyper-rectangles of the form 
\begin{equation}
{\mathcal M^{k}} =\{\vec p\;:\;r_{i}^{(k)}\leq p_{i}\leq s_{i}^{(k)},\;\forall\;i=1, \ldots, L\}\triangleq [\vec r^{(k)},\vec s^{(k)}]\;.
\end{equation} 
In each hyper-rectangle $\mathcal M^k$, we require an upper-bound $\beta(\mathcal M^k)$ of \cref{opt} on $\mathcal M^k$, i.e., a power vector $\widetilde{\vec p}^{(k)}\in \mathcal M^k$ feasible for \cref{opt} with \cgls{wsee} greater than or equal to the \gls{wsee} achieved by any other power vector in $\mathcal M^k$. Such a bound over the box $\mathcal M^k = [ \vec r^{(k)}, \vec s^{(k)}]$ is
\ifhbonecolumn
	\begin{align}
		\MoveEqLeft \underset{\vec p\in\mathcal M^{k}}{\text{max}}\ \sum_{i=1}^L w_i \frac{\log\left( 1 + \frac{\alpha_i p_i}{1 + \sum_{j\neq i} \beta_{i,j} p_j} \right)}{\mu_i p_i + P_{c,i}}
		\le \sum_{i=1}^L w_i\; \underset{\vec p\in\mathcal M^{k}}{\text{max}} \frac{\log\left( 1 + \frac{\alpha_i p_i}{1 + \sum_{j\neq i} \beta_{i,j} p_j} \right)}{\mu_i p_i + P_{c,i}}\notag \\
		\label{Eq:SumEEBound}
		&= \sum_{i=1}^L w_i\; \underset{r_{i}^{(k)} \le p_i \le s_i^{(k)}}{\text{max}} \underbrace{\frac{\log\left( 1 + \frac{\alpha_i p_i}{1 + \sum_{j\neq i} \beta_{i,j} r^{(k)}_j} \right)}{\mu_i p_i + P_{c,i}}}_{\coloneqq \overline{\text{EE}}_i(p_i, \mathcal M^{k})} \eqqcolon \beta(\mathcal M^k)
	\end{align}
\else
	\begin{align}
		\MoveEqLeft \underset{\vec p\in\mathcal M^{k}}{\text{max}}\ \sum_{i=1}^L w_i \frac{\log\left( 1 + \frac{\alpha_i p_i}{1 + \sum_{j\neq i} \beta_{i,j} p_j} \right)}{\mu_i p_i + P_{c,i}}\notag \\
		&\le \sum_{i=1}^L w_i\; \underset{\vec p\in\mathcal M^{k}}{\text{max}} \frac{\log\left( 1 + \frac{\alpha_i p_i}{1 + \sum_{j\neq i} \beta_{i,j} p_j} \right)}{\mu_i p_i + P_{c,i}}\notag \\
		&= \sum_{i=1}^L w_i\; \underset{r_{i}^{(k)} \le p_i \le s_i^{(k)}}{\text{max}} \underbrace{\frac{\log\left( 1 + \frac{\alpha_i p_i}{1 + \sum_{j\neq i} \beta_{i,j} r^{(k)}_j} \right)}{\mu_i p_i + P_{c,i}}}_{\coloneqq \overline{\text{EE}}_i(p_i, \mathcal M^{k})}\notag\\
		\label{Eq:SumEEBound}
		&\eqqcolon \beta(\mathcal M^k)
	\end{align}
\fi
where the last step is due to $\text{EE}_i$ being decreasing in $p_j$ for all $j\neq i$. Thus, we  compute the bound by maximizing each $\text{EE}_i$ with respect to $\vec p$ over $\mathcal M^k $. This provides an excellent accuracy-complexity trade-off and leads to fast convergence, as confirmed by the numerical analysis reported in \cref{sec:numeval}. At the same time, the bound can be computed with reasonably low computational complexity. Indeed, $\overline{\text{EE}}_i(p_i, \mathcal M^k)$ is a strictly pseudo-concave function of $p_{i}$, being the ratio of a strictly concave over an affine function \cite{ZapNow15}. Thus, its global maximizer is obtained as the unique zero of its derivative, namely the unique solution of the equation:
\begin{equation} \label{eq:deriv}
	\frac {\alpha_i (\mu_i p_i + P_{c,i})}{1 + \sum_{j\neq i} \beta_{i,j} r^{(k)}_j + \alpha_i p_i} = \mu_i \ln\bigg(1 + \frac{\alpha_i p_i}{1 + \sum_{j\neq i} \beta_{i,j} r^{(k)}_j}\bigg),
\end{equation}
which is denoted by $\widetilde p_i^{(k)}\!$. 
Equation~\cref{eq:deriv} can be solved numerically with any root finding algorithm, e.g. with Newton-Raphson's or Halley's method. However, due to $\frac{\mathrm{d}}{\mathrm{d} p_i} \overline{\text{EE}}_i$ approaching zero quickly as $p_i \rightarrow \infty$ these methods might suffer from numerical problems. Instead, a more stable numerical solution of \cref{eq:deriv} is computed as
\begin{equation}\label{Eq:PowerIterate}
	\hat p^{(k)}_i = \frac{1}{\widetilde\alpha_i} \left( \frac{\frac{\widetilde\alpha_i}{\mu_i} P_{c,i} - 1}{W_0\left(\left(\frac{\widetilde\alpha_i}{\mu_i} P_{c,i} - 1\right) e^{-1}\right)} - 1 \right),
\end{equation}
where $\widetilde\alpha_i = \frac{\alpha_i}{1 + \sum_{j\neq i} \beta_{i,j} r^{(k)}_j}$ and $W_0(\cdot)$ is the principal branch of the Lambert $W$ function. The point $\hat p^{(k)}_i$ might be outside $\mathcal M^k$. Due to the pseudo-concavity of $\overline{\text{EE}}_i(p_i, \mathcal M^k)$, the optimal solution is
\begin{equation}
	\widetilde p^{(k)}_i =
	\begin{cases}
		r_i^{(k)} &\text{if}\ p^{(k)}_i \le r_i^{(k)} \\
		\hat p^{(k)}_i &\text{if}\ \hat r_i^{(k)} < p^{(k)}_i < s_i^{(k)} \\
		s_i^{(k)} &\text{otherwise}
	\end{cases}
\end{equation}

This bound is tight at $\vec r^{(k)}$, i.e., $\text{EE}_i(\vec r^{(k)}) = \overline{\text{EE}}_i(r^{(k)}_i, \mathcal M^k)$, and the bounding procedure 
generates a point $\widetilde{\vec p}^{(k)}_i$ that is often in the interior of $\mathcal M^k$. This allows the use of an adaptive bisection rule that drives the bound directly towards an objective value. Instead, typical exhaustive bisection strives to reduce the size of each partition element towards a singleton which results in much slower convergence. In particular, in each iteration $k$ the hyper-rectangle $\mathcal M^k = [\vec r^{(k)}, \vec s^{(k)}]$ with the best bound is selected and then bisected via $(\vec v^{(k)}, j_k)$ where 
\begin{equation} \label{eq:bisectv}
	\vec v^{(k)} = \frac{1}{2} (\widetilde{\vec p}^{(k)} + \vec r^{(k)}), \quad j_k = \argmax_j \abs{\widetilde{p}^{(k)}_j - r^{(k)}_j}.
\end{equation}
The partition sets are given by the subrectangles $\mathcal M_-^k$ and $\mathcal M_+^k$ determined by the hyperplane $p_{j_k} = v_{j_k}$ as
\begin{equation}
	\begin{aligned}
		\mathcal M^k_- &= \{ \vec x \,|\, r^{(k)}_{j_k} \le x_{j_k} \le v^{(k)}_{j_k},\ r^{(k)}_i \le x_i \le s^{(k)}_i\ (i\neq j_k) \} \\
		\mathcal M^k_+ &= \{ \vec x \,|\, v^{(k)}_{j_k} \le x_{j_k} \le s^{(k)}_{j_k},\ r^{(k)}_i \le x_i \le s^{(k)}_i\ (i\neq j_k) \}.
	\end{aligned}
	\label{eq:bisect}
\end{equation}

The final procedure is stated in \cref{alg:bb}. The set $\mathscr R_k$ holds all hyper-rectangles to be examined and $\gamma$ is the \cgls{cbv}. In lines~\ref{l1}--\ref{l2}, the box $\mathcal M^k$ with the best bound is selected and bisected. Later, in line~\ref{l5}, this hyper-rectangle is removed from $\mathscr R_k$. Each of the resulting subrectangles is examined in lines~\ref{l3}--\ref{l4}. If its bound is not better than the \cgls{cbv} (plus the absolute tolerance $\varepsilon$), it is ignored. Otherwise, it is added to $\mathscr R_k$ in line~\ref{l5}. Then, the algorithm selects a feasible point from $\mathcal M^k$ and updates the \cgls{cbs} if the objective value for this point is better than the \cgls{cbv}. When  $\mathscr R_k$ is empty, there are no further hyper-rectangles to examine and the problem is solved.

\begin{algorithm}[t]
	\caption{Global optimal solution of \cref{opt}}\label{alg:bb}
	\begin{algorithmic}[1]
		\small
		\State Initialize $\mathscr R_0 = \{[\vec 0, \vec P]\}$, $\gamma = -\infty$, $k = 0$
		\Statex
		\Repeat
			\State Select $\mathcal M^k \in \argmax\{\beta(\mathcal M) \,|\, \mathcal M \in \mathscr R_k \}$. \label{l1}
			\State Bisect $\mathcal M^k$ via $(\vec v^k, j_k)$ with $(\vec v^k, j_k)$ as in \cref{eq:bisectv} and
			\State Let $\mathscr P_k = \{ \mathcal M_-^k, \mathcal M_+^k \}$ with $\mathcal M_-^k$, $\mathcal M_+^k$  as in \cref{eq:bisect}. \label{l2}
			\Statex
			\ForAll {$\mathcal M \in \mathscr P_k$} \label{l3}
				\If {$\beta(\mathcal M) > \gamma+\varepsilon$} \label{l6}
					\State Add $\mathcal M$ to $\mathscr S_k$.
					\State Let $\vec r$ such that $\mathcal M = [\vec r, \vec s]$.
					\If {$f(\vec r) > \gamma$}
						\State $\bar{\vec p} \gets \vec r$
						\State $\gamma \gets f(\vec r)$
					\EndIf
				\EndIf
			\EndFor \label{l4}
			\Statex
			\State $\mathscr R_{k+1} \gets \mathscr S_k \cup \mathscr R_k\setminus\{\mathcal M^k\}$ \label{l5}
			\State $k \gets k+1$
		\Until{$\mathscr R_k = \emptyset$}
		\Statex
		\State\Return $\bar{\vec p}$ as the optimal solution
	\end{algorithmic}
\end{algorithm}

Convergence of \cref{alg:bb} is stated formally below.
\begin{proposition} \label{thm:conv}
	For every $\varepsilon > 0$, \cref{alg:bb} converges in a finite number of iterations towards a point with objective value within an $\varepsilon$-region of the global optimal value of \cref{opt}.
\end{proposition}

\begin{IEEEproof}
	First, observe that the branching procedure does not generate any box containing infeasible points. Hence, no feasibility checks are necessary during \cref{alg:bb}.
	
	After each iteration, $\mathscr R_k$ contains all boxes that might hold a better solution than the \cgls{cbs} $\bar{\vec p}$. If the algorithm terminates, then all boxes $\mathcal M$ generated from $\mathscr R_k$ after the last update of $\gamma$ had a bound $\beta(\mathcal M) \le \gamma + \varepsilon$. Since $\bar{\vec p}$ is feasible and satisfies $f(\bar{\vec p}) > \beta(\mathcal M) - \varepsilon \ge \underset{\vec p \in \mathcal M}{\text{max}} f(\vec p) - \varepsilon$ for every $\mathcal M$, $\bar{\vec p}$ is a global $\varepsilon$-optimal solution. Thus, it remains to show that the algorithm is finite, i.e., that the termination criterion $\mathscr R_k = \emptyset$ occurs at some point.
By virtue of \cite[Prop.~6.2]{Tuy2016}, this is the case if both points $\widetilde{\vec p}^{(k)}, \vec r^{(k)}$ in \cref{eq:bisectv} are in $\mathcal M^k$, $\widetilde{\vec p}^{(k)}$ is feasible, and the bounding procedure satisfies
\begin{equation}\label{eq:adaptiveconv}
f\big(\vec r^{(k_v)}\big) - \beta\big(\mathcal M^{k_v}\big) = o\big(\big\Vert\widetilde{\vec p}^{(k)} - \vec r^{(k)}\big\Vert\big).
\end{equation}
Since, by \cite[Thm.~6.4]{Tuy2016}, there exists a subsequence $\{k_v\}_v$ such that $\vec r^{(k_v)}$ and $\widetilde{\vec p}^{(k_v)}$ approach a common limit $\vec x$ as $v\to\infty$, it holds that
\ifhbonecolumn
\begin{equation*}
	f(\vec r^{(k_v)}) - \beta(\mathcal M^{k_v})
	= \sum_{i=1}^L w_i \overline{\text{EE}}_i(r^{(k_v)}_i, \mathcal M^{k_v}) - \sum_{i=1}^L w_i \overline{\text{EE}}_i(\widetilde p^{(k_v)}_i, \mathcal M^{k_v})
	\rightarrow 0,
\end{equation*}
\else
\begin{multline*}
	f(\vec r^{(k_v)}) - \beta(\mathcal M^{k_v})\\
	= \sum_{i=1}^L w_i \overline{\text{EE}}_i(r^{(k_v)}_i, \mathcal M^{k_v}) - \sum_{i=1}^L w_i \overline{\text{EE}}_i(\widetilde p^{(k_v)}_i, \mathcal M^{k_v})
	\rightarrow 0,
\end{multline*}
\fi
and, thus, $\beta(\mathcal M^{k_v}) \to f(\vec x)$.
This point is feasible for \cref{opt} and, hence, $\beta(\mathcal M^{k_v}) \ge f(\vec x)$. Therefore, $\beta(\mathcal M^{k_v}) \to f(\vec x^*)$ with $\vec x^*$ the optimal solution of \cref{opt} as $v \to\infty$.
\end{IEEEproof}
The following remarks are in order.
\begin{remark}
	Inspecting \cref{alg:bb}, it can be seen that each step towards the computation of the optimal power allocation is continuous with respect to the channel parameters $\{\alpha_{i},\beta_{i,j}\}_{i,j}$, with the exception of \eqref{Eq:PowerIterate}, which has a discontinuity in $\alpha_{i}=0$. Therefore, in order to claim the continuity of the map \eqref{Eq:F1}, which will be required by \cref{Prop:UAP}, it is necessary to assume that $\alpha_{i}$ is bounded below by a strictly positive quantity, i.e., $\alpha_{i}> \omega_{i}$, for some $\omega_{i}>0$ and $i=1, \ldots, L$. This assumption does not seem restrictive for any practical system, recalling that $\alpha_{i}$ is the channel-to-noise ratio between the $i$-th \gls{ue} and the associated \gls{bs}.
\end{remark}
\begin{remark}
If it is desired to use a relative tolerance instead of an absolute tolerance, it is sufficient to replace ``$\beta(\mathcal M) > \gamma + \varepsilon$'' by ``$\beta(\mathcal M) > (1 + \varepsilon) \gamma$'' in line~\ref{l6} of \cref{alg:bb}.
\end{remark}

\subsection{Application to other Energy Efficiency Metrics}\label{Sec:Extensions}
This section explicitly shows how to apply the proposed framework to other performance metrics. It was already observed in \cref{Sec:Motivation} that the \gls{wsee} generalizes the system weighted sum-rate defined in \eqref{Eq:WSR}, and thus the proposed approach naturally applies to the maximization of the weighted sum-rate with $\widetilde{\vec p}^{(k)} = \vec s^{(k)}$ as can be easily seen from \cref{Eq:SumEEBound}. Instead, the application of our framework to other \cgls{ee} metrics requires more effort. Specifically, we will consider the WPEE, WMEE, and GEE functions, as defined in \eqref{Eq:WPEE}, \eqref{Eq:WMEE}, and \eqref{Eq:GEE}.

\subsubsection{WPEE and WMEE maximization}
Let us consider the WPEE function. The main issue regarding the applicability of the proposed \cgls{bb} procedure is the derivation of a bound for the WPEE over the generic hyper-rectangle ${\mathcal M}^k$, that is tight, satisfies \cref{eq:adaptiveconv}, and is simple to maximize. To this end, observe that we can write
\ifhbonecolumn
	\begin{align}
		\MoveEqLeft \max_{\vec p\in\mathcal{M}^{k}}\; \prod_{i=1}^{L}\frac{\log\left(1+\frac{\alpha_{i}p_{i}}{1+\sum_{j\neq i}\beta_{i,j}p_{j}}\right)}{\mu_{i}p_{i}+P_{c,i}}
		\le\prod_{i=1}^{L}\max_{\vec p\in\mathcal{M}^{k}}\;\frac{\log\left(1+\frac{\alpha_{i}p_{i}}{1+\sum_{j\neq i}\beta_{i,j}p_{j}}\right)}{\mu_{i}p_{i}+P_{c,i}}\notag\\
	\label{Eq:ProdEEBound}
		&=\prod_{i=1}^{L}\max_{r_{i}^{(k)}\leq p_{i}\leq s_{i}^{(k)}} \frac{\log\left(1+\frac{\alpha_{i}p_{i}}{1+\sum_{j\neq i}\beta_{i,j}r_{j}^{(k)}}\right)}{\mu_{i}p_{i}+P_{c,i}}
		\eqqcolon \beta(\mathcal M^k).
	\end{align}
\else
	\begin{align}
		\MoveEqLeft \max_{\vec p\in\mathcal{M}^{k}}\; \prod_{i=1}^{L}\frac{\log\left(1+\frac{\alpha_{i}p_{i}}{1+\sum_{j\neq i}\beta_{i,j}p_{j}}\right)}{\mu_{i}p_{i}+P_{c,i}} \notag\\
		&\le\prod_{i=1}^{L}\max_{\vec p\in\mathcal{M}^{k}}\;\frac{\log\left(1+\frac{\alpha_{i}p_{i}}{1+\sum_{j\neq i}\beta_{i,j}p_{j}}\right)}{\mu_{i}p_{i}+P_{c,i}}\notag\\
	\label{Eq:ProdEEBound}
		&=\prod_{i=1}^{L}\max_{r_{i}^{(k)}\leq p_{i}\leq s_{i}^{(k)}} \frac{\log\left(1+\frac{\alpha_{i}p_{i}}{1+\sum_{j\neq i}\beta_{i,j}r_{j}^{(k)}}\right)}{\mu_{i}p_{i}+P_{c,i}} \\
		&\eqqcolon \beta(\mathcal M^k).
	\end{align}
\fi
Note that, for any $i=1, \ldots, L$, the upper-bound $\beta(\mathcal M^k)$ in \eqref{Eq:ProdEEBound} coincides with that obtained for the \gls{wsee} function in \eqref{Eq:SumEEBound}. Thus, the same approach used for \gls{wsee} maximization applies also to the maximization of the WPEE. 

Moreover, it can be seen that the same bounding technique applies also to the WMEE function. Indeed, all steps above can be made also if the product of the \glspl{ee} is replaced by the minimum of the \glspl{ee}. Indeed, both are increasing functions of the individual \glspl{ee} and no differentiability assumption is required by the proposed \cgls{bb} procedure. 

\subsubsection{GEE Maximization}
The \gls{gee} function does not explicitly depend on the individual \glspl{ee}, which slightly complicates the bounding technique. Nevertheless, a similar approach applies. Defining $P_{c}=\sum_{i=1}^{L}P_{c,i}$, we have
\ifhbonecolumn
	\begin{align}
		\MoveEqLeft \max_{\vec p\in\mathcal{M}^{k}}\; \frac{\sum_{i=1}^{L}\log\left(1+\frac{\alpha_{i}p_{i}}{1+\sum_{j\neq i}\beta_{i,j}p_{j}}\right)}{P_{c}+\sum_{i=1}^{L}\mu_{i}p_{i}}
		\le\sum_{i=1}^{L}\max_{\vec p\in\mathcal{M}^{k}}\frac{\log\left(1+\frac{\alpha_{i}p_{i}}{1+\sum_{j\neq i}\beta_{i,j}p_{j}}\right)}{P_{c}+\sum_{i=1}^{L}\mu_{i}p_{i}}\notag\\
		\label{Eq:BoundGEE}
		&=\sum_{i=1}^{L}\max_{s_{i}^{(k)}\leq p_{i}\leq r_{i}^{(k)}} \frac{\log\left(1+\frac{\alpha_{i}p_{i}}{1+\sum_{j\neq i}\beta_{i,j}r_{j}^{(k)}}\right)}{\mu_{i}p_{i}+P_{c}+\sum_{j\neq i}\mu_{j}r_{j}^{(k)}}
		\eqqcolon \beta(\mathcal M^k),
	\end{align}
\else
	\begin{align}
		\MoveEqLeft \max_{\vec p\in\mathcal{M}^{k}}\; \frac{\sum_{i=1}^{L}\log\left(1+\frac{\alpha_{i}p_{i}}{1+\sum_{j\neq i}\beta_{i,j}p_{j}}\right)}{P_{c}+\sum_{i=1}^{L}\mu_{i}p_{i}} \notag\\
		&\le\sum_{i=1}^{L}\max_{\vec p\in\mathcal{M}^{k}}\frac{\log\left(1+\frac{\alpha_{i}p_{i}}{1+\sum_{j\neq i}\beta_{i,j}p_{j}}\right)}{P_{c}+\sum_{i=1}^{L}\mu_{i}p_{i}}\notag\\
		\label{Eq:BoundGEE}
		&=\sum_{i=1}^{L}\max_{s_{i}^{(k)}\leq p_{i}\leq r_{i}^{(k)}} \frac{\log\left(1+\frac{\alpha_{i}p_{i}}{1+\sum_{j\neq i}\beta_{i,j}r_{j}^{(k)}}\right)}{\mu_{i}p_{i}+P_{c}+\sum_{j\neq i}\mu_{j}r_{j}^{(k)}} \\
		&\eqqcolon \beta(\mathcal M^k),
	\end{align}
\fi
where the last step is due to $\frac{\log\left(1+\frac{\alpha_{i}p_{i}}{1+\sum_{j\neq i}\beta_{i,j}p_{j}}\right)}{P_{c}+\sum_{\ell=1}^{L}\mu_{\ell}p_{\ell}}$ being decreasing\footnote{This can be verified easily from the first-order derivative.} in $p_j$ for all $j\neq i$.
Although slightly different from that obtained for the \gls{wsee}, WPEE, and WMEE, the bound in \eqref{Eq:BoundGEE} is formally equivalent to \eqref{Eq:SumEEBound} as a function of $p_{i}$. Thus, the same maximization procedure applies to \gls{gee} maximization, too. 

\section{First-order Optimal Power Control}\label{Sec:SEQ}
This section is devoted to developing a power control algorithm with guaranteed convergence to a first-order optimal point of Problem~\eqref{opt}. Besides providing an alternative power control approach with polynomial complexity, the method developed here also provides a theoretically solid benchmark for other approaches. The algorithm proposed in this section is inspired by the successive pseudo-concave framework from \cite{Yang2017}, which tackles the maximization of a function $f$ by maximizing a sequence of approximate functions $\{\widetilde{f}_{t}\}_{t}$ fulfilling the following properties for all $j$:
\begin{enumerate}
	\item\label{a1} $\widetilde f_{t}(\vec p; \vec p^{(t)})$ is pseudo-concave in $\vec p$ for any feasible $\vec p^{(t)}$.
	\item\label{a2} $\widetilde f_{t}(\vec p; \vec p^{(t)})$ is continuously differentiable in $\vec p$ for any feasible $\vec p^{(t)}$ and continuous in $\vec p^{(t)}$ for any feasible $\vec p$.
	\item\label{a3} $\nabla_{\vec p} \widetilde f_{t}(\vec p^{(t)}; \vec p^{(t)}) = \nabla_{\vec p} f(\vec p^{(t)})$.
	\item\label{a4} $\widetilde f_{t}(\vec p; \vec p^{(t)})$ has a non-empty set of maximizers in the feasible set. 
	\item\label{a5} Given any convergent subsequence $\{\vec p^{(t)}\}_{t\in\mathcal T}$ where $\mathcal T\subseteq\{1, 2, \ldots\}$, the sequence $\{\argmax_{\vec p\in\mathcal P}\ \widetilde f_{t}(\vec p; \vec p^{(t)})\}_{t\in\mathcal T}$ is bounded.
\end{enumerate}
It is seen that the approximate functions are parametrized by $\vec p^{(t)}$ and the properties above need to hold for any feasible $\vec p^{(t)}$. In practice, $\vec p^{(t)}$ is updated according to a specific rule after each iteration, as will be explained in the sequel. 

Under the assumptions above, \cite{Yang2017} shows that every limit point of the sequence ${\vec p}_{t}^{*}$ of maximizers of $\widetilde f_{t}$ with respect to $\vec p$, converges to a first-order optimal point for the original problem of maximizing $f$. Moreover, due to the first property above, for all $t$,  the maximization of $\widetilde f_{t}$ can be accomplished with polynomial complexity by standard optimization methods \cite{Boyd2004}.\footnote{Recall that pseudo-concave functions are differentiable by definition and every stationary point is a global maximizer.} Thus, the crucial point when employing the above framework is about finding suitable approximate functions $\{\widetilde f_{t}\}_{t}$ that fulfill all above assumptions. Next, it is shown how this can be accomplished for the \gls{wsee} maximization Problem~\eqref{opt}.

To elaborate, recalling \eqref{Eq:Rate}, and for a given point $\vec p^{(t)}$, we propose the following approximation for EE$_{i}$ in \eqref{eq:eei}, 
\begin{multline}\label{Eq:EEapprox}
w_{i}\text{EE}_{i}\approx\widetilde{\text{EE}}_{i}(\vec p; \vec p^{(t)}) =
	w_i \frac{R_i(p_i, \vec p^{(t)}_{-i})}{\mu_i p^{(t)}_i + P_{c,i}}
	+
	(p_i - p^{(t)}_i) \\
	\cdot\left( w_i \frac{-\mu_i R_i(\vec p^{(t)})}{(\mu_i p^{(t)}_i + P_{c,i})^2}
	+
	\sum_{j\neq i} w_j
	\frac{\frac{\partial}{\partial p_i} R_j(\vec p^{(t)})}{\mu_i p^{(t)}_i + P_{c,i}}
	\right),
\end{multline}
wherein $R_i(p_i, \vec p^{(t)}_{-i})$ denotes the $i$-th user's rate as a function of the $i$-th user's power $p_{i}$, while all other powers are fixed to $\vec p^{(t)}_{-i}=\{p^{(t)}_{j}\}_{j\neq i}$. 

In the Appendix it is shown that the approximation \eqref{Eq:EEapprox} fulfills all five properties required by the successive pseudo-concave approximation framework. Here, we remark that, for all $i=1, \ldots, L$, the approximate function in \eqref{Eq:EEapprox} is constructed in such a way to be concave with respect to $p_{i}$, since the $i$-th user's rate $R_{i}$ is concave in $p_{i}$, while the other terms in the \gls{ee} have been linearized around $\vec p = \vec p^{(t)}$. Replacing each \gls{ee} in the \gls{wsee} maximization problem \eqref{opt} by \eqref{Eq:EEapprox}, we obtain the approximate problem:
\begin{optprob}
	& \underset{\vec p}{\text{max}}
	&& \sum_{i=1}^L  \widetilde{\text{EE}}_{i}(\vec p; \vec p^{(t)})\\
	& \text{s.\,t.}
	&& 0 \le p_i \le P_i,\quad\text{for all } i = 1, 2, \ldots, L,
	\label{opt_approx1}
\end{optprob}
which is a concave maximization problem and can therefore be solved in polynomial time using standard convex optimization tools \cite{Nesterov1994}. Moreover, since $\widetilde{\text{EE}}_{i}$ depends only on $p_{i}$ for all $i$, Problem~\eqref{opt_approx1} can be decoupled over the users, and each transmit power $p_{i}$ can be optimized separately by solving the scalar problem:
\begin{optprob}
	& \underset{p_{i}}{\text{max}}
	&&\widetilde{\text{EE}}_{i}(p_{i}; \vec p^{(t)})\\
	& \text{s.\,t.}
	&& 0 \le p_i \le P_i,
	\label{opt_approx2}
\end{optprob}
Thus, by the successive pseudo-concave optimization framework, the original Problem~\eqref{opt} is tackled by solving a sequence of problems of the form of \eqref{opt_approx1}, updating the point $\vec p^{(t)}$ after each iteration according to the formula
\begin{equation}
\vec p^{(t+1)} = \vec p^{(t)} + \gamma^{(t)} (\mathds B\vec p^{(t)} - \vec p^{(t)})\;,
\end{equation}
with $\mathds B\vec p^t$ an optimal solution of \cref{opt_approx1} and $\gamma^t = \beta^{m_t}$ to be determined by the Armijo rule, where $m^t$ is the smallest nonnegative integer such that
\ifhbonecolumn
\begin{equation*}
	f(\vec p^{(t)} + \beta^{m_t} (\mathds B\vec p^{(t)} - \vec p^{(t)})) \ge f(\vec p^{(t)}) + \alpha \beta^{m_t} \nabla f(\vec x^{(t)})^T (\mathds B\vec p^{(t)} - \vec p^{(t)})
\end{equation*}
\else
\begin{multline*}
	f(\vec p^{(t)} + \beta^{m_t} (\mathds B\vec p^{(t)} - \vec p^{(t)})) \\\ge f(\vec p^{(t)}) + \alpha \beta^{m_t} \nabla f(\vec x^{(t)})^T (\mathds B\vec p^{(t)} - \vec p^{(t)})
\end{multline*}
\fi
with $0<\alpha<1$ and $0<\beta<1$ being scalar constants.\footnote{Please refer to \cite[Sec.\ 1.2.1]{Bertsekas1999} for more details on the Armijo rule and how to choose $\alpha$ and $\beta$ properly.}

The overall procedure is summarized in \cref{alg:spca}, whose convergence is formally stated below.
\begin{proposition} \label{thm:spca}
	Any limit point of $\{\vec p^{(t)}\}_t$ obtained by \cref{alg:spca} is a stationary point of \cref{opt}. 
\end{proposition}

\begin{IEEEproof}
	Please refer to the Appendix.
\end{IEEEproof}

\begin{algorithm}
\caption{Successive convex approximation algorithm}\label{alg:spca}
\begin{algorithmic}[1]
	\small
	\State Initialize $t = 0$, $\vec p^{(0)} \in [\vec 0, \vec P]$, $\alpha, \beta\in(0, 1)$.
	\Repeat
	\State $\mathds B_i \vec p^{(t)} \gets \argmax_{0\le p_i \le P_i} \widetilde r_i(p_i; \vec p^{(t)}),\quad i = 1, 2, \dots, L.$
	\State $\gamma^{(t)} \gets 1$
	\ifhbonecolumn
		\While {$f(\vec p^{(t)} + \gamma^{(t)} (\mathds B\vec p^{(t)} - \vec p^{(t)})) < f(\vec p^{(t)}) + \alpha \gamma^{(t)} \nabla f(\vec p^{(t)})^T (\mathds B\vec p^{(t)} - \vec p^{(t)})$}
	\else
		\parWhile {$f(\vec p^{(t)} + \gamma^{(t)} (\mathds B\vec p^{(t)} - \vec p^{(t)})) < f(\vec p^{(t)}) + {}$ \flushright $\alpha \gamma^{(t)} \nabla f(\vec p^{(t)})^T (\mathds B\vec p^{(t)} - \vec p^{(t)})$}
	\fi
			\State $\gamma^{(t)} \gets \beta\gamma^{(t)}$
	\ifhbonecolumn
		\EndWhile
	\else
		\EndparWhile
	\fi
		\State $\vec p^{(t+1)} \gets \vec p^{(t)} + \gamma^{(t)} (\mathds B\vec p^{(t)} - \vec p^{(t)})$
		\State $t \gets t+1$
	\Until{convergence.}
\end{algorithmic}
\end{algorithm}

\section{\gls{ann}-Based Power Control}\label{Sec:ANN}
While the global optimization method from \cref{Sec:BB} is aimed at providing an efficient way to solve offline Problem~\eqref{opt}, this section proposes an approach that is able to tackle Problem~\eqref{opt} with a complexity that is amenable to an online implementation, i.e., with a complexity that enables to update the optimal power control vector with the same frequency as the variations of the fading channel realizations. This will be achieved by merging the global optimization method from \cref{Sec:BB} with an \gls{ann}-based procedure. The main idea is based on reformulating the optimization problem \eqref{opt} as the problem of determining the map 
\begin{equation}\label{Eq:F1}
{\cal F}:{\bf a}=( \alpha_{i}, \beta_{i,j}, P_{i} )_{i,j}\in\mathbb{R}^{L(L+1)} \mapsto {{\boldsymbol {p}}}^{*}\in\mathbb{R}^{L},
\end{equation}
with ${\boldsymbol {p}}^{*}$ the optimal power allocation corresponding to ${\bf a}$. Thus, \glspl{ann} can be used to learn the map \eqref{Eq:F1}, since \glspl{ann}, and in particular feedforward neural networks with fully-connected layers, are universal function approximators \cite{Leshno1993}. 

\subsection{Network Architecture}
Before proceeding further, let us first briefly introduce the architecture of the considered \gls{ann}. We employ a fully-connected feedforward network which takes as input a realization of the parameter vector ${\bf a}$, producing as output a power allocation vector $\hat{\vec p}$, which estimates the optimal power allocation vector ${\bf p}^{*}$ corresponding to ${\bf a}$. Between the input and output layer, $K$ hidden layers are present. For all $k=1, \ldots, K+1$, the $k$-th layer has $N_{k}$ neurons, with neuron $n$ computing 
\begin{equation}\label{Eq:TransferFunction}
	\zeta_{k}(n)=f_{n,k}\!\left(\vec\gamma_{n,k}^{T}\vec\zeta_{k-1}+\delta_{n,k}\right)
\end{equation}
wherein $\vec\zeta_{k} = (\zeta_k(1), \dots, \zeta_k(N_{k+1}))$ denotes the $N_{k+1}\times 1$ output vector of layer $k$, $\vec\gamma_{n,k}\in\mathbb{R}^{N_{k-1}}$ and $\delta_{n,k}\in\mathbb{R}$ are neuron-dependent weights and bias terms, respectively, while $f_{n,k}$ is the activation function\footnote{In principle, any function can be considered as activation function, even though widely accepted choices are sigmoidal functions, rectified linear units (ReLU), and generalized ReLU functions. The specific choices considered in this work are discussed in \cref{sec:numeval}.} of neuron $n$ in layer $k$.
\begin{figure}
	\centering
	\tikzsetnextfilename{annModel}
	\begin{tikzpicture}[xscale=.8, yscale=.6]
		\begin{scope}[every node/.style={circle, draw,inner sep=4}]
			\node (inp1) at (0,0) {};
			\node (inpN) at (0,-3) {};

			\node (h11) at (2,.5) {};
			\node (h12) at (2,-.5) {};
			\node (h1N) at (2,-3.5) {};

			\node (h21) at (4,0) {};
			\node (h22) at (4,-1) {};
			\node (h2N) at (4,-3) {};

			\node (h31) at (6,-.5) {};
			\node (h3N) at (6,-2.5) {};

			\node (out1) at (8,0) {};
			\node (outN) at (8,-3) {};
		\end{scope}

		\begin{scope}[-latex]
			\draw (inp1) -- (h11);
			\draw (inp1) -- (h12);
			\draw (inp1) -- (h1N);
			\draw (inpN) -- (h11);
			\draw (inpN) -- (h12);
			\draw (inpN) -- (h1N);
			
			\draw (h11) -- (h21);
			\draw (h11) -- (h22);
			\draw (h11) -- (h2N);
			\draw (h12) -- (h21);
			\draw (h12) -- (h22);
			\draw (h12) -- (h2N);
			\draw (h1N) -- (h21);
			\draw (h1N) -- (h22);
			\draw (h1N) -- (h2N);

			\draw (h21) -- (h31);
			\draw (h21) -- (h3N);
			\draw (h22) -- (h31);
			\draw (h22) -- (h3N);
			\draw (h2N) -- (h31);
			\draw (h2N) -- (h3N);

			\draw (h31) -- (out1);
			\draw (h31) -- (outN);
			\draw (h3N) -- (out1);
			\draw (h3N) -- (outN);

			\draw ($(inp1.west) + (-25pt,0)$) -- (inp1);
			\draw ($(inpN.west) + (-25pt,0)$) -- (inpN);

			\draw (out1) -- ($(out1.east) + (25pt,0)$);
			\draw (outN) -- ($(outN.east) + (25pt,0)$);
		\end{scope}

		\begin{scope}[densely dashed]
			\draw ($(h11.north -| h11.west) + (-10pt,5pt)$) coordinate (hRectNW)
				rectangle ($(h1N.south -| h3N.east) + (10pt,-5pt)$) coordinate (hRectSE);

			\draw ($(h11.north -| inp1.west) + (-10pt,5pt)$) coordinate (inRectNW)
				rectangle ($(h1N.south -| inpN.east) + (10pt,-5pt)$) coordinate (inRectSE);

			\draw ($(h11.north -| out1.west) + (-10pt,5pt)$) coordinate (outRectNW)
				rectangle ($(h1N.south -| outN.east) + (10pt,-5pt)$) coordinate (outRectSE);
		\end{scope}

		\node [anchor=south] at ($(hRectNW)!.5!(hRectNW -| hRectSE)$) {Hidden Layers};
		\node [anchor=south] at ($(inRectNW)!.5!(inRectNW -| inRectSE)$) {Input Layer};
		\node [anchor=south] at ($(outRectNW)!.5!(outRectNW -| outRectSE)$) {Output Layer};

		\begin{scope}[decoration={markings, mark=at position .5 with {\fill circle (.9pt); \fill (5pt,0) circle (.9pt); \fill (-5pt,0) circle (.9pt);}}]
			\path [postaction=decorate] (inp1) -- (inpN);
			\path [postaction=decorate] (h12) -- (h1N);
			\path [postaction=decorate] (h22) -- (h2N);
			\path [postaction=decorate] (h31) -- (h3N);
			\path [postaction=decorate] (out1) -- (outN);
		\end{scope}
	\end{tikzpicture}
  \caption{General scheme of a deep feedforward \gls{ann} with fully-connected layers.}
  \label{Fig:ANN_Scheme}
\end{figure}
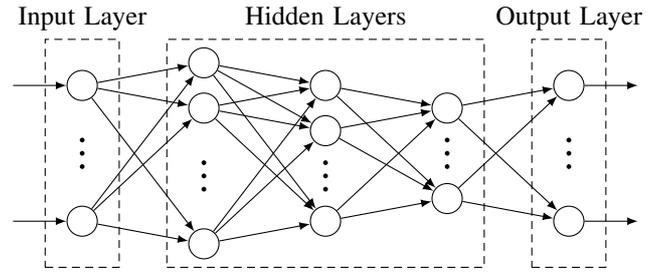
Apparently, each neuron performs quite simple operations. Nevertheless, combining the processing of   multiple neurons, \glspl{ann} can perform very complex tasks and obtain an overall input-output map that emulates virtually any function. Formally, the following universal approximation result holds \cite[Theorem 1]{Leshno1993}.
\begin{proposition}\label{Prop:UAP}
Consider a single layer of an \gls{ann} with $n\times 1$ input vector $\vec \xi$, $m\times n$ weight matrix $\widetilde{\vec \Gamma}$, $m\times 1$ bias vector $\widetilde{\vec \delta}$, and activation functions $\vec f=[f_{1}, \ldots, f_{m}]$. The set of all input-output maps that can be obtained by the \gls{ann} is
\begin{equation}
\mathcal A_{n}=\left\{\vec f\big(\widetilde{\vec \Gamma}\vec \xi +\widetilde{\vec \delta}\big) \,\big|\; \widetilde{\vec \Gamma}\in\mathbb{R}^{n\times m}, \widetilde{\vec \delta}\in\mathbb{R}^{m}\right\}\;.
\end{equation}
Then, $\mathcal A_{n}$ is dense in the set of continuous functions if and only if $\vec f$ is not an algebraic polynomial. 
\end{proposition}

\medskip
\Cref{Prop:UAP} formally proves that the input-output relationship of an \gls{ann} can emulate any continuous map.\footnote{The continuity of \eqref{Eq:F1} has been analyzed in \cref{Sec:BB}.} In addition to \cref{Prop:UAP}, \cite{Barron1993} provides bounds for the number of neurons to be used to obtain a given approximation accuracy. However, although of great theoretical importance, neither of these results is constructive, in the sense that they do not provide any guidance as to the topology of the \gls{ann} to use and how to configure the weights and biases to achieve a desired approximation accuracy. In practice, it has been empirically shown that deep architectures, i.e., \gls{ann} with multiple layers, tend to require less neurons \cite[Sec. 6.4.1]{Bengio2016} to achieve the same level of accuracy, which motivates us to employ a deep \gls{ann}. Moreover, it is intuitively clear that the number of neurons to employ increases with the size of the problem, i.e., with the dimension of the domain and co-domain of the map to estimate. For the case of \eqref{Eq:F1}, it is needed to estimate a map from an $L(L+1)$-dimensional space to an $L$-dimensional space.

\subsection{Training Procedure, Normalization, and Data Augmentation}\label{Sec:Data-Augmentation}
At this point, the problem remains of how to tune the weights and biases to reliably estimate \eqref{Eq:F1}. To this end, the weights $\vec\Gamma=\{\vec\gamma_{n,k}\}_{n,k}$, and the biases $\vec\delta=\{\delta_{n,k}\}_{n,k}$ are adjusted in a supervised learning fashion by training the \gls{ann}. This requires the use of a training set, i.e., a set of the form
$\{ (\vec a_n, \vec p_n^{*}) \,|\, n = 1, \dots, N_T \}$ with $N_T$ training tuples $(\vec a_n, \vec p_n^{*})$, wherein $\vec p_n^{*}$ is the optimal power allocation vector corresponding to $\vec a_n$. In other words, the training set contains examples of desired power allocation vectors corresponding to some possible configurations of system parameters $\vec a_n$. By exploiting these examples, the \gls{ann} learns to predict the power allocation also for new realizations of $\vec a_n$ that are not contained in the training set. Mathematically speaking, the training process consists of adjusting the weights and biases of the \gls{ann} in order to minimize the loss between actual and desired output, namely considering the problem:
\begin{equation}\label{Eq:TrainingMin}
	\displaystyle\min_{\vec\Gamma, \vec\delta}\;\frac{1}{N_{T}}\sum_{n=1}^{N_{T}}{\cal L}(\hat{\vec p}_{n}(\vec\Gamma, \vec\delta), {\boldsymbol {p}}_{n}^{*})
\end{equation}
with ${\cal L}(\cdot,\cdot)$ being any suitable measure of the error incurred when the actual output of the \gls{ann} corresponding to the $n$-th training input is $\hat{\vec p}_{n}$, while the desired output was ${\bf p}_{n}^{*}$. A widely-used error measure is the squared error $\norm{\hat{\vec p}_{n}(\vec\Gamma, \vec\delta) - {\boldsymbol {p}}_{n}^{*}}^2$ \cite{Bengio2016} which is also employed here. The minimization of \eqref{Eq:TrainingMin} can be tackled by state-of-the-art, off-the-shelf stochastic gradient descent methods specifically developed for training \glspl{ann} \cite{Bengio2016}, and therefore will not be discussed here. Instead, it is interesting to note that the learning process can be simplified by normalizing the transmit powers before running the 
stochastic gradient descent training algorithm. Specifically, applying the variable change $p_i \to \tilde p_i P_i$, for all $i = 1, \dots, L$, we normalize the transmit power to lie in the interval $[0, 1]$, which  
leads to the following equivalent reformulation of Problem~\eqref{opt}
\begin{optprob}
	& \underset{\tilde{\vec p}}{\text{max}}
	&& \sum_{i=1}^L  w_{i}\frac{\log\left( 1 + \frac{\tilde{\alpha}_i \tilde{p}_i}{1 + \sum_{j\neq i} \tilde{\beta}_{i,j} \tilde{p}_j} \right)}{\mu_{i} P_{i} \tilde{p}_i + P_{c,i}} \\
	& \text{s.\,t.}
	&& 0 \le \tilde{p}_i \le 1,\quad\text{for all } i = 1, 2, \ldots, L,
	\label{opt2}
\end{optprob}
wherein $\tilde{\alpha}_{i}=\alpha_{i} P_{i}$, $\tilde{\beta}_{i}=\beta_{i} P_{i}$, $\tilde{\mu}_{i}=\mu_{i} P_{i}$, for all $i=1, \ldots, L$. Then, the normalized training set then is ${\mathcal S}_{T} = \{ (\tilde{\vec a}_n, \tilde{\vec p}_n^{*}) \,|\, n = 1, \dots, N_T \}$ with parameter vector $\tilde{\vec a} = ( \tilde\alpha_{i}, \tilde\beta_{i,j}, P_{i} )_{i,j}$. The advantage of this reformulation is that, despite the values $P_{1}, \ldots, P_{L}$, the transmit powers are always in the set $[0,1]$. Intuitively, this simplifies the dependence of the optimal power allocation on the maximum power constraints, thereby making it easier for the \gls{ann} to grasp the optimal power allocation structure as a function of the maximum power constraints. 

When doing this, the use of realistic numbers for the receive noise power and propagation channels might lead to coefficients $\{\tilde{\alpha}_{i}, \tilde{\beta}_{i,j}\}_{i,j}$ with quite a large magnitude, which often cause numerical problems to the stochastic gradient descent training algorithm. We have observed that this issue is solved by expressing the parameter vectors $\tilde{\vec a}$ in the training set in logarithmic units rather than in a linear scale. A similar problem occurs for the output powers, which in some cases might be close to zero due to the normalization by $P_{\text{max}}$. This issue is also resolved by expressing the output powers in logarithmic scale. On the other hand, logarithms cause numerical problems when the optimal transmit powers are very close to zero. In order to avoid this issue, a suitable approach is to clip logarithmic values approaching $-\infty$ at $-M$ for $M > 0$. Thus, the considered normalized training set is
\begin{equation*}
{\mathcal S}_{T} = \{ (\log_{10} \tilde{\vec a}_n, \max\{-M, \log_{10} \tilde{\vec p}_n^{*}\}) \,|\, n = 1, \dots, N_T \}.
\end{equation*}

Additionally, we can use the fact that the problem is invariant under permutation of the users for data augmentation, i.e., we can increase the size of the training set $\mathcal{S}_{T}$ during training. In order to do this, the rows and columns of a channel matrix (and the corresponding power allocations) from the training set can be permuted to generate a new training sample. In mathematical terms, given a permutation $\sigma$ of the index set $\mathcal{I}=\left\{1, 2, \dots{}, L\right\}$, the elements of the new matrix $\tilde{H}$ are given as $\tilde{h}_{i,j}=h_{\sigma(i),\sigma(j)}$ by permuting the indices of a channel matrix $H$ from the training set.
In each training step, a new random permutation is generated and used to permute the training samples and corresponding labels. Therefore, the \gls{ann} automatically learns the invariance of the problem against permutation of the users. 

An important aspect for good training performance is the choice of the output layer where the proposed \cgls{ann} deploys a linear activation function. This
seems to contrast with the fact that the transmit powers need to be constrained in the interval $[0, 1]$. However, enforcing this constraint directly in the output activation function might mislead the \gls{ann}. Indeed, it could lead to low training  errors simply thanks to the use of cut-off levels in the activation function, instead of being the result of proper adjustment of the hidden layer weights and biases. In this case, the \gls{ann} would not be able to learn that the training and validation errors are acceptable only because the desired power level is close to either 1 or 0, and the clipping at the output layer provides by construction such a power level, regardless of the configuration adopted in the hidden layers. Instead, a linear output activation function allows the \gls{ann} to learn whether the present configuration of weights and biases is truly leading to a small error. At the end of the training phase, the
output variables are clipped to the interval $[0, 1]$.

After the training phase, all weights and biases of the \gls{ann} are configured and the \gls{ann} essentially provides a closed-form estimate of the map \eqref{Eq:F1}. Indeed, once the weights and biases have been set, the input-output relationship of the \gls{ann} can be written in closed-form as the composition of the affine combinations and activation functions of the neurons in the \gls{ann}. This effectively provides a closed-form expression for the map \eqref{Eq:F1}, within an approximation accuracy that can be made small at will by properly designing and training the \gls{ann}. Thus, as ${\bf a}$ changes due to channel fading, the corresponding power allocation can be obtained without having to solve Problem \cref{opt} again, but simply computing the output of the \gls{ann} when the input is the new realization of ${\bf a}$. This grants a large complexity reduction during the online operation of the method, as compared to other approaches that employ iterative methods where several convex problems need to be solved for each instance of the channel realizations. This point is analyzed in more detail in the  next section. 

\subsection{Computational Complexity}\label{Sec:Comlpexity}
The main advantage of the proposed \gls{ann}-based method is that it allows performing most of the computations towards solving \eqref{opt} offline, leaving only a few operations to be executed online, i.e., only a few operations need to be repeated when the system channel realizations vary, while most of the computations need to be performed only sporadically. This is in contrast to available online power control methods based on the traditional use of optimization theory, which need to be run from scratch every time one or more system channel realizations have changed. To elaborate, the complexity of the proposed \gls{ann}-based power allocation can be divided into an online and an offline complexity, as explained next:
\begin{itemize}
	\item[(a)] \textbf{Online complexity.} This is the complexity that is required to use the trained \gls{ann} for the online computation of the power allocation vector. As discussed below, this is the complexity that is incurred during the online operation of the method, i.e., when the trained \gls{ann} is being used to output the optimal power control policy following the variations of the channel fading realizations. 
\item[(b)] \textbf{Offline complexity.} This is the complexity that is required to build the training set and to implement the training procedure. As discussed below, these tasks can be executed at a much longer time-scale than that with which the channel fading realizations change. 
\end{itemize}

\textbf{Online phase.} It requires computing the output $\zeta_{k}(n)$ of each neuron in the \gls{ann}, moving from the first layer to the output layer, which in turn requires $\sum_{k=1}^{K+1}N_{k-1}N_{k}$ real multiplications,\footnote{The complexity related to additions is neglected as it is much smaller than that required for multiplications.} and evaluating $\sum_{k=1}^{K+1}N_{k}$ scalar activation functions $f_{n,k}$. Despite being typically non-linear, the activation functions are elementary functions whose computation  does not pose any significant computational issue. Thus, obtaining the output of the trained \gls{ann} for any given input vector entails a negligible complexity, since it requires only the computation of a forward propagation of the trained \gls{ann}. For this reason, the online complexity of the proposed \gls{ann}-based method is much lower than the complexity of the first-order optimal method from \cref{Sec:SEQ}, which instead requires solving convex problems in each iteration.

\textbf{Offline phase.} It requires the generation of the training set and its use to train the \gls{ann}. Among these two tasks, the most complex is the generation of the training set, since the execution of the training algorithm is conveniently performed by off-the-shelf stochastic gradient descent algorithms, which ensure a fast convergence. Moreover gradient computation is performed by the backpropagation algorithm, which further reduces the computational complexity \cite{Bengio2016}. Instead, generating the training set requires actually solving the NP-complete Problem \cref{opt} $N_{T}$ times, i.e., for $N_{T}$ different realizations of the system parameter vector ${\bf a}$. At a first sight, this might seem to defeat the purpose of using the proposed \gls{ann}-based approach, but actually this is not the case for three main reasons:
\begin{itemize}
\item The whole training phase (including both generation and use of the training set) can be performed  \emph{offline}. Thus, a much higher complexity can be afforded, and it is not needed to complete the training process within the channels coherence time. 
\item The training set can be updated at a \emph{much longer time-scale} than the channels coherence time. In other words, the training set can be updated sporadically compared with the frequency with which Problem \cref{opt} should be solved if traditional optimization approaches were used.
\item Despite the first two points, it can be argued that Problem \cref{opt} is NP-complete, and thus generating a large training set appears a daunting task even if it can be performed offline. However, the global optimization method that is proposed in \cref{Sec:BB} eases this issue, making it possible  to globally solve \cref{opt} in practical wireless networks, with a complexity that is affordable for offline implementations. 
\end{itemize}

Finally, we explicitly observe that, as anticipated, the proposed \gls{ann}-based approach is not restricted to the maximization of the \gls{wsee}. Indeed, any power allocation problem can be cast as in \eqref{Eq:F1} and the \cgls{bb} method developed in Section~\ref{Sec:BB} is not limited to the maximization of the \gls{wsee}, but encompasses all major energy-efficient metrics, as addressed in detail in \cref{Sec:Extensions}. 

\section{Numerical Evaluation} \label{sec:numeval}
We consider the uplink of a wireless interference network. At first, we consider that $L=4$ single-antenna \cglspl{ue} are placed in a square area with edge \unit[2]{km} and communicate with 4 access points placed at coordinates \unit[$(0.5, 0.5)$]{km}, \unit[$(0.5, 1.5)$]{km}, \unit[$(1.5, 0.5)$]{km}, \unit[$(1.5, 1.5)$]{km}, and equipped with $n_{R}=2$ antennas each. The path-loss is modeled following \cite{PathLossModel}, with carrier frequency \unit[1.8]{GHz} and power decay factor equal to 4.5, while fast fading terms are modeled as realizations of zero-mean, unit-variance circularly symmetric complex Gaussian random variables. The circuit power consumption and power amplifier inefficiency terms are equal to $P_{c,i} = \unit[1]{W}$ and $\mu_{i} = 4$ for all $i=1, \ldots, L$, respectively. The noise power at each receiver is generated as $\sigma^{2}=F{\cal N}_{0}B$, wherein $F=\unit[3]{dB}$ is the receiver noise figure, $B=\unit[180]{kHz}$ is the communication bandwidth, and ${\cal N}_{0}=\unit[-174]{dBm/Hz}$ is the noise spectral density. All users have the same maximum transmit powers $P_{1} =  \cdots = P_{L}=P_{\text{max}}$.

The proposed \gls{ann}-based solution of Problem~\eqref{opt} is implemented through a feedforward \gls{ann} with $K+1$ fully-connected layers, with the $K=5$ hidden layers having 128, 64, 32, 16, 8 neurons, respectively. In order to generate the training set, Problem~\eqref{opt} needs to be solved for different realizations of the vector $\tilde{\vec a} = (\tilde{\alpha}_i, \tilde{\beta}_{i,j}, P_{\text{max}})_{i,j}$. The data is converted to logarithmic units as explained in \cref{Sec:Data-Augmentation} with clipping at $-M = -20$.\footnote{Note that, although using a logarithmic scale, the transmit powers are not expressed in dBW, since the logarithmic values are not multiplied by 10. Thus $-M=-20$, corresponds to \unit[-200]{dBW}.}
As for the activation functions, ReLU and its generalizations are the most widely used choice. Our experiments verify that they also perform well in this application. Specifically,
the first hidden layer has an exponential linear unit (ELU) activation, motivated by the need to compensate for the logarithmic conversion in the training set. This choice, together with the logarithmic normalization of the data set, has proven itself essential for good training performance.
The other hidden layers alternate ReLU and ELU activation functions while the output layer deploys a linear activation function (cf.~\cref{Sec:Data-Augmentation}).

\subsection{Training Performance}
The \cgls{ann} is implemented in Keras~2.2.4\cite{keras} with TensorFlow~1.12.0\cite{tensorflow} as backend.
Training is performed on a Nvidia GeForce GTX 1080 Ti over 500 epochs with batches of size 128 and shuffling of the training data before each epoch. Initialization is performed by Keras with default parameters, i.e., Glorot uniform initialization \cite{glorot10} for the kernel and zero biases. The optimization problem \cref{Eq:TrainingMin} is solved by the Adam optimizer with Nesterov momentum \cite{sutskever13}, initialized by Keras default parameters, with the squared error as the loss function in \eqref{Eq:TrainingMin}. Source code and data sets are available online \cite{github}.

The training set is generated from 2000 \cgls{iid} realizations of \cglspl{ue}' positions and propagation channels. Users are randomly placed in the service area and channels are generated according to the channel model described above. Each \cgls{ue} $i$ is associated to the access point towards which it enjoys the strongest effective channel $\alpha_{i}$. 
For each channel realization, we apply \cref{alg:bb} to solve \cref{opt} for $P_{\text{max}} = -30, \ldots, 20\,\textrm{dB}$ in \unit[1]{dB} steps with relative tolerance $\varepsilon = 0.01$. This yields a training set of 102,000 samples.

Besides the training set, also a validation set and a test set are required. The validation set is used during training to estimate the generalization performance of the \cgls{ann}, i.e., the performance on a data set the \cgls{ann} was not trained on.
The validation loss is the central metric for hyperparameter tuning, i.e., choosing all parameters of the \cgls{ann} other than weights and biases (e.g. number of layers, activation functions, batch size). Since during this process information about the validation set leaks into the \gls{ann} model, another set for the final testing of the \gls{ann} is required, the test set. It is essential that the test set is never used during training and tuning of the \cgls{ann} \cite{Chollet2017}. The validation and test sets have been independently generated from 200 and 10,000 \cgls{iid} channel realizations, respectively, with the same procedure used for the training set. This results in 10,200 samples for the validation set, i.e., 10\% of the training set, and 510,000 samples for the test set. The final performance to be shown in \cref{Sec:Testing} will be averaged over the test set samples. Thus, using a test set based on 10,000 channel scenarios means that the performance presented in \cref{Sec:Testing} is what is obtained by using the trained \gls{ann} for 10,000 channel coherence times. This confirms that the training phase needs to be performed only sporadically.

Considering training, validation, and test sets, 622,200 data samples were generated, which required solving the NP-complete problem \cref{opt} 622,200 times. This has been accomplished by the newly proposed \gls{bb} method developed in \cref{alg:bb}, which has been implemented in C++ employing the  Intel\textregistered~MKL, OpenMP, and the Lambert $W$ library published at \url{https://github.com/CzB404/lambert_w}. Computing the complete data set (622,200 samples including training, validation, and test sets) took 8.4 CPU hours on Intel Haswell nodes with Xeon E5-2680 v3 CPUs running at \unit[2.50]{GHz}. The mean and median times per sample are \unit[48.7]{ms} and \unit[4.8]{ms}, respectively, which shows the effectiveness of the proposed \cref{alg:bb}, and in turn supports the argument that the offline generation of a suitable training set for the proposed \gls{ann}-based power control method is quite affordable. 

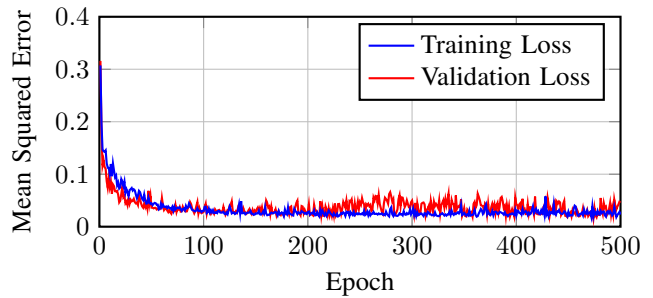
\begin{figure}
\centering
\begin{tikzpicture}
	\begin{axis} [
			thick,
			xlabel={Epoch},
			ylabel={Mean Squared Error},
			ylabel near ticks,
			grid=major,
			no markers,
			xmin = 0,
			xmax = 500,
			ymin = 0,
			ymax = .4,
			legend pos=north east,
			legend cell align=left,
			width=\axisdefaultwidth,
			height=.6*\axisdefaultheight,
			reverse legend,
		]

		\pgfplotstableread[col sep=comma]{training_4users_rerun.dat}\tbl

		\addplot+[red] table[y=val_loss] {\tbl};
		\addlegendentry{Validation Loss};
		\addplot+[blue] table[y=loss] {\tbl};
		\addlegendentry{Training Loss};
	\end{axis}
\end{tikzpicture}
\vspace{-2ex}
\caption{Training and validation loss.}
\label{fig:training}
\vspace{-2ex}
\end{figure}

Due to the random initialization, shuffling of the training data, and the inherent randomness of the optimizer, the weights and biases of the \cgls{ann} are realizations of a random process. Thus, all performance results reported for the \cgls{ann} are averaged over 10 realizations of the network obtained by training the \cgls{ann} \emph{on the same training set} with different initialization of the underlying random number generator.\footnote{Note that this is not equivalent to \emph{model ensembling} \cite[Sect.~7.3.3]{Chollet2017} or \emph{bagging} \cite[Sect.~7.1]{Bengio2016}.}
The average training and validation losses for the final \cgls{ann} are shown in \cref{fig:training}. It can be observed that both errors quickly approach a small value of the same order of magnitude as the tolerance of \cref{alg:bb}. Moreover, neither of the losses increases over time which leads to the conclusion, that the adopted training procedure fits the training data well, without underfitting or overfitting. 

\tikzsetnextfilename{loss}
\tikzpicturedependsonfile{training_4users_rerun.dat}

\subsection{Testing Performance}\label{Sec:Testing}
\begin{figure}
\centering
\begin{tikzpicture}
	\begin{axis} [
			thick,
			xlabel={$P_{\text{max}}$ [dBW]},
			ylabel={WSEE [Mbit/Joule]},
			ylabel near ticks,
			grid=major,
			no markers,
			xmin = -30,
			xmax = 20,
			ymin = 0,
			legend pos=south east,
			legend cell align=left,
			cycle list name=long,
			width=\axisdefaultwidth,
			height=.8*\axisdefaultheight,
			legend columns = 3,
			legend style = {
				at = {(.5,1.03)},
				anchor=south,
				/tikz/every even column/.append style={column sep=0.25cm}
			},
		]

		\pgfplotstableread[col sep=comma]{verification_4users_rerun.dat}\tbl

		\addplot+[smooth] table[y=opt] {\tbl};
		\addlegendentry{Optimal};

		\addplot+[smooth] table[y=ANN] {\tbl};
		\addlegendentry{ANN};

		\addplot+[smooth] table[y=SCA] {\tbl};
		\addlegendentry{SCA};

		\addplot+[smooth] table[y=SCAos] {\tbl};
		\addlegendentry{SCAos};

		\addplot+[smooth] table[y=max] {\tbl};
		\addlegendentry{Max.~Power};

		\addplot+[smooth] table[y=best] {\tbl};
		\addlegendentry{Best only};

	\end{axis}
\end{tikzpicture}
\vspace{-2ex}
\caption{Performance on the test set compared to the global optimal solution, first-order optimal solutions, and fixed power allocations.}
\label{fig:testset}
\vspace{-2ex}
\end{figure}
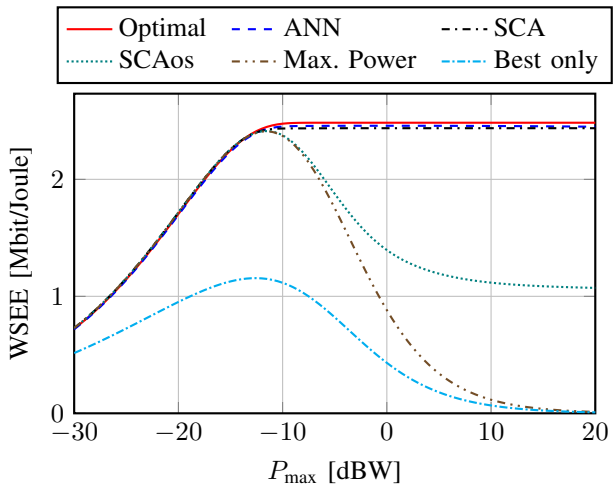

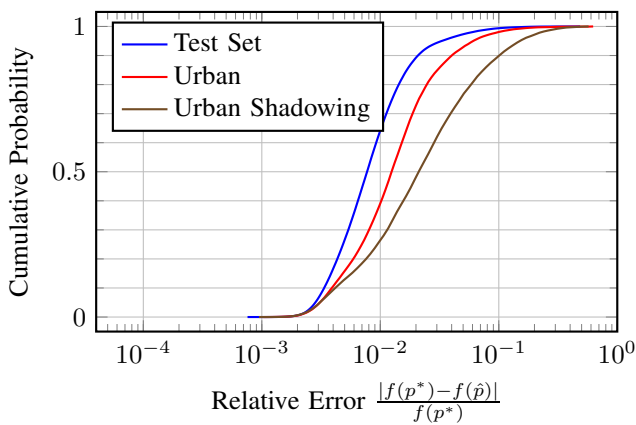
\begin{figure}
\centering
\begin{tikzpicture}
	\begin{semilogxaxis} [
			thick,
			xlabel={Relative Error $\frac{\abs{f(p^*) - f(\hat p)}}{f(p^*)}$},
			ylabel={Cumulative Probability},
			ylabel near ticks,
			grid=major,
			minor y tick num = 4,
			yminorgrids = true,
			no markers,
			xmin = .00004,
			xmax = 1,
			ymin = -0.05,
			ymax = 1.05,
			legend pos=north west,
			legend cell align=left,
			width=\axisdefaultwidth,
			height=.8*\axisdefaultheight,
		]

		\addplot table [col sep=comma] {verification_relerr_4users_rerun.dat};
		\addlegendentry{Test Set};

		\addplot table [col sep=comma, x=x, y=y] {generalization_relerr_hataUrban_noSF_4users_rerun.dat};
		\addlegendentry{Urban};

		\addplot table [col sep=comma, x=x, y=y] {generalization_relerr_hataUrban_4users_rerun.dat};
		\addlegendentry{Urban Shadowing};
	\end{semilogxaxis}
\end{tikzpicture}
\vspace{-2ex}
\caption{Empirical \cgls{cdf} of the relative approximation error.}
\label{fig:cdf}
\vspace{-2ex}
\end{figure}

The average performance of the final \cgls{ann} on the test set is reported in \cref{fig:testset}. Recall that this test set is never used during training and, thus, the \cgls{ann} has no information about it except for its statistical properties gathered from the training set (and, possibly, the validation set due to hyperparameter tuning).
It can be seen from \cref{fig:testset} that the gap to the optimal value is virtually non-existent which is confirmed by the empirical \cgls{cdf} of the relative approximation error displayed in \cref{fig:cdf}. Its mean and median values are 0.0133 and 0.00739 respectively.

\tikzsetnextfilename{WSEE}
\tikzpicturedependsonfile{verification_4users_rerun.dat}

\tikzsetnextfilename{WSEEcdf}
\tikzpicturedependsonfile{verification_relerr_4users_rerun.dat}
\tikzpicturedependsonfile{generalization_relerr_hataUrban_4users_rerun.dat}
\tikzpicturedependsonfile{generalization_relerr_hataUrban_noSF_4users_rerun.dat}

In addition to near-optimal performance and low computational complexity, the proposed \gls{ann}-based approach also outperforms several baseline approaches. Specifically, we have included a comparison with the following benchmarks:
\begin{itemize}
\item \textbf{SCAos:} The first-order optimal method based on the sequential convex approximation method developed in \cref{Sec:SEQ}. For each value of $P_{\text{max}}$, the algorithm initializes the transmit power to $p_{i}=P_{\text{max}}$, for all $i=1, \ldots, L$. 
\item \textbf{SCA:} This is again the first-order optimal method based on sequential convex approximation developed in \cref{Sec:SEQ}, but in this case a double-initialization approach is used. Specifically, at $P_{\text{max}} = \unit[-30]{dBW}$ once again maximum power initialization is used. However, for all values of $P_{\text{max}} > \unit[-30]{dBW}$, the algorithm is run twice, once with the maximum power initialization, and once initializing the transmit powers with the optimal solution obtained for the previous $P_{\text{max}}$ value. Then, the power allocation achieving the better \gls{wsee} value is retained.
\item \textbf{Max. Power:} All \cglspl{ue} transmit with $p_{i}=P_{\text{max}}$, for all $i=1, \ldots, L$. This strategy is known to perform well in interference networks for low $P_{\text{max}}$ values.
\item \textbf{Best only:} Only one \cgls{ue} is allowed to transmit, specifically that with the best effective channel. This approach is motivated for high $P_{\text{max}}$ values, as a naive way of nulling out multi-user interference.
\end{itemize}
The results show that the proposed \gls{ann}-based approach outperforms all other suboptimal schemes. The only exception is the SCA approach which shows similar (but still worse) performance. However, as described above, this method relies on a sophisticated initialization rule, which requires to solve the \gls{wsee} maximization problem twice and for the complete range of $P_{\text{max}}$ values. This is clearly not suitable for obtaining a "one-shot" solution, i.e., when the \gls{wsee} needs to be maximized only for one specific value of $P_{\text{max}}$, as is required for online resource allocation. Moreover, it requires some calibration depending on the channel statistics since it performs well provided the $P_{\text{max}}$ range starts sufficiently far away from the \gls{wsee} saturation region, i.e., the range of $P_{\text{max}}$ values for which the \gls{wsee} keeps constant, starting from $P_{\text{max}} \approx \unit[-10]{dBW}$ in \cref{fig:testset}. Thus, the SCA approach has a quite higher complexity than the \gls{ann}-based method, but, despite this, it  performs slightly worse. In conclusion, we can argue that the \cgls{ann} approach is much better suited to online power allocation than state-of-the-art approaches, including \cref{alg:spca}.

\subsection{Resilience against Channel Modeling Mismatches}
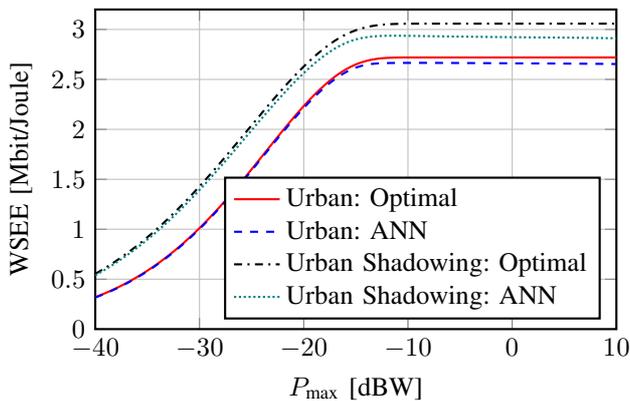
\begin{figure}
\centering
\begin{tikzpicture}
	\begin{axis} [
			thick,
			xlabel={$P_{\text{max}}$ [dBW]},
			ylabel={WSEE [Mbit/Joule]},
			ylabel near ticks,
			grid=major,
			ytick = {0,.5,...,4},
			no markers,
			xmin = -40,
			xmax = 10,
			ymin = 0,
			ymax = 3.2,
			legend pos=south east,
			legend cell align=left,
			cycle list name=long,
			width=\axisdefaultwidth,
			height=.8*\axisdefaultheight,
		]

		\pgfplotstableread[col sep=comma]{generalization_4users_rerun.dat}\tbl

		\addplot+[smooth] table[y=hataUrban_noSF_opt] {\tbl};
		\addlegendentry{Urban: Optimal};

		\addplot+[smooth] table[y=hataUrban_noSF_ANN] {\tbl};
		\addlegendentry{Urban: ANN};

		\addplot+[smooth] table[y=hataUrban_opt] {\tbl};
		\addlegendentry{Urban Shadowing: Optimal};

		\addplot+[smooth] table[y=hataUrban_ANN] {\tbl};
		\addlegendentry{Urban Shadowing: ANN};
	\end{axis}
\end{tikzpicture}
\vspace{-2ex}
\caption{Performance on different channel distributions: Hata-COST231 Urban propagation model with and without \unit[8]{dB} shadowing.}
\label{fig:otherchan}
\vspace{-2ex}
\end{figure}

Previous results consider a test set whose samples are independently generated from the training and validation sets, but following the same statistical distribution. Instead, now we analyze how robust the \cgls{ann} performance is to changes in the channel statistics. To this end, in the following we consider a new test set, whose samples are generated according to a different statistical distribution. Specifically, we generate path-loss effects according to the Hata-COST231 propagation model \cite{HataCost231,Rappaport2002} for urban (non-metropolitan) areas with carrier frequency \unit[1.9]{GHz} and base station height \unit[30]{m}. However, we do not repeat the training based on the new channel model, but instead use the same \gls{ann} trained as described above. 
Remarkably, as indicated by the prediction performance reported in \cref{fig:otherchan} and verified by the distribution of the relative error in \cref{fig:cdf} under the label ``Urban,'' the performance  degrades only slightly compared to the case in which the test and training samples come from the same distribution.

Next, we further modify the channel generation procedure of the test set, by also introducing log-normal shadowing \cite{Rappaport2002} with \unit[8]{dB} standard deviation. It can be observed from \cref{fig:cdf} that the median relative error increases by roughly half an order of magnitude. Given that the underlying channel distribution is quite different from the trained channel model, the performance can still be considered good. This is also verified by the \cgls{wsee} performance shown in \cref{fig:otherchan}. Indeed, the performance is still better than with SCAos. Based on these observations, we conclude that training the \cgls{ann} on synthetic data based on simple channel models is quite robust and performs well also in more sophisticated channels scenarios. Of course, the performance tends to degrade as the mismatch between the training and test set distributions increases.

\tikzsetnextfilename{WSEEgeneralization}
\tikzpicturedependsonfile{generalization_4users_rerun.dat}

\subsection{Performance of an \cgls{ann} with Reduced Size}
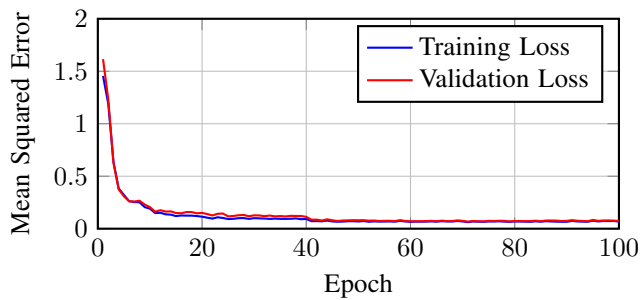
\begin{figure}
\centering
\begin{tikzpicture}
	\begin{axis} [
			thick,
			xlabel={Epoch},
			ylabel={Mean Squared Error},
			ylabel near ticks,
			grid=major,
			no markers,
			xmin = 0,
			xmax = 100,
			ymin = 0,
			ymax = 2,
			legend pos=north east,
			legend cell align=left,
			width=\axisdefaultwidth,
			height=.6*\axisdefaultheight,
		]

		\pgfplotstableread[col sep=comma]{training_16.dat}\tbl

		\addplot table[y=loss] {\tbl};
		\addlegendentry{Training Loss};

		\addplot table[y=val_loss] {\tbl};
		\addlegendentry{Validation Loss};
	\end{axis}
\end{tikzpicture}
\vspace{-2ex}
\caption{Training and validation loss of the smaller \cgls{ann}.}
\label{fig:training16}
\vspace{-2ex}
\end{figure}

\begin{figure}
\centering
\begin{tikzpicture}
	\begin{semilogxaxis} [
			thick,
			xlabel={Relative Error $\frac{\abs{f(p^*) - f(\hat p)}}{f(p^*)}$},
			ylabel={Cumulative Probability},
			ylabel near ticks,
			grid=major,
			minor y tick num = 4,
			yminorgrids = true,
			no markers,
			xmin = .0001,
			xmax = 1,
			ymin = -0.05,
			ymax = 1.05,
			legend pos=north west,
			legend cell align=left,
			width=\axisdefaultwidth,
			height=.8*\axisdefaultheight,
		]

		\addplot table [col sep=comma] {verification_relerr_16.dat};
		\addlegendentry{Test Set};

		\addplot table [col sep=comma, x=x, y=y] {generalization_relerr_hataUrban_noSF_16.dat};
		\addlegendentry{Urban};

		\addplot table [col sep=comma, x=x, y=y] {generalization_relerr_hataUrban_16.dat};
		\addlegendentry{Urban Shadowing};
	\end{semilogxaxis}
\end{tikzpicture}
\vspace{-2ex}
\caption{Empirical \cgls{cdf} of the relative approximation error made by the smaller \cgls{ann}.}
\label{fig:cdf16}
\vspace{-2ex}
\end{figure}
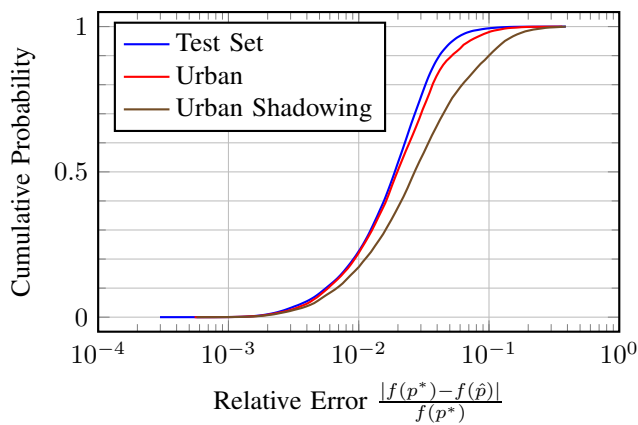

\begin{figure}
\centering
\begin{tikzpicture}
	\begin{axis} [
			thick,
			xlabel={$P_{\text{max}}$ [dBW]},
			ylabel={WSEE [Mbit/Joule]},
			ylabel near ticks,
			grid=major,
			ytick = {0,.5,...,4},
			no markers,
			xmin = -40,
			xmax = 10,
			ymin = 0,
			ymax = 3.2,
			legend pos=south east,
			legend cell align=left,
			cycle list name=long,
			width=\axisdefaultwidth,
			height=.8*\axisdefaultheight,
		]

		\pgfplotstableread[col sep=comma]{generalization_16.dat}\tbl

		\addplot+[smooth] table[y=hataUrban_noSF_opt] {\tbl};
		\addlegendentry{Urban: Optimal};

		\addplot+[smooth] table[y=hataUrban_noSF_ANN] {\tbl};
		\addlegendentry{Urban: ANN};

		\addplot+[smooth] table[y=hataUrban_opt] {\tbl};
		\addlegendentry{Urban Shadowing: Optimal};

		\addplot+[smooth] table[y=hataUrban_ANN] {\tbl};
		\addlegendentry{Urban Shadowing: ANN};
	\end{axis}
\end{tikzpicture}
\vspace{-2ex}
\caption{Performance of the smaller \cgls{ann} on different channel distributions: Hata-COST231 Urban propagation model with and without \unit[8]{dB} shadowing.}
\label{fig:otherchan16}
\vspace{-2ex}
\end{figure}
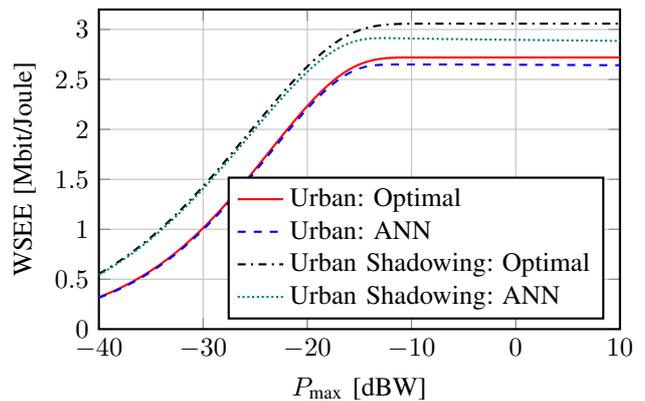

Next, we evaluate the performance of a much smaller \cgls{ann} trained with the same data as before. Specifically, we only consider 2 hidden layers having 16 and 8 neurons, respectively, with activation functions ELU and ReLU and no permutation of the training data. This further reduces the computational complexity for online resource allocation. Again, the output layer has 4 nodes and a linear activation function. Training is performed in batches of size 128. While the counterpart of \cref{fig:testset} looks identical (and is, therefore, not reproduced), the difference between the two \cglspl{ann} is best studied from the training loss in \cref{fig:training16} and the distribution of the relative error in \cref{fig:cdf16}. First, observe from \cref{fig:training16} that the training and validation losses stall on a value clearly greater than those of the original (larger) \cgls{ann}. The \cgls{cdf} of the relative error reflects this as well, being shifted significantly to the right. Still, the mean error on the test set is so small that no difference would be observed in terms of achieved \gls{wsee} value.
Instead, the performance on the test set generated from the Hata-COST231 Urban model with shadowing differs noticeably from the original \cgls{ann} as can be seen from \cref{fig:otherchan16}. However, the performance can still be considered good. Thus, although the smaller \cgls{ann} has worse performance than the original, the practical implications are limited and its reduced complexity might be worth the downsides.

\tikzpicturedependsonfile{training_16.dat}

\tikzsetnextfilename{WSEEcdf16}
\tikzpicturedependsonfile{verification_relerr_16.dat}
\tikzpicturedependsonfile{generalization_relerr_hataUrban_16.dat}
\tikzpicturedependsonfile{generalization_relerr_hataUrban_noSF_16.dat}

\tikzsetnextfilename{WSEEgeneralization16}
\tikzpicturedependsonfile{generalization_16.dat}

\subsection{Increasing the Number of Users}
In the previous sections, a scenario with $L=4$ users has been considered. In this section, we increase the number of users to demonstrate that resource allocation with \cglspl{ann} can be scaled to a higher  number of users and further showcase the performance of the proposed global optimization algorithm. In particular, consider the same scenario as before but with $L = 7$ \cglspl{ue} and $n_R = 4$ antennas per access point.

To account for the increased randomness in the data set, the sizes of the \cgls{ann} and data sets need to be increased. An \cgls{ann} with 
nine hidden layers having (1024, 4096, 1024, 512, 256, \dots{}, 16) nodes, respectively, ELU activation functions on the first and last layers, and ReLU in all others layers has shown the best performance in our numerical experiments. The training set has been generated from 6000 \cgls{iid} channel realizations, while the validation and test sets were generated from 600 and 1000 channels, respectively. This results in a total of 387,600 samples, labelled with \cref{alg:bb} with mean and median computation times of \unit[260.6]{s} and \unit[24.1]{s} per sample, respectively.

\begin{figure}
	\centering
	\begin{tikzpicture}
	\begin{axis} [
	thick,
	xlabel={$P_{\text{max}}$ [dBW]},
	ylabel={WSEE [Mbit/Joule]},
	ylabel near ticks,
	grid=major,
	no markers,
	xmin = -30,
	xmax = 20,
	ymin = 0,
	legend pos=south east,
	legend cell align=left,
	cycle list name=long,
	width=\axisdefaultwidth,
	height=.8*\axisdefaultheight,
	legend columns = 3,
	legend style = {
		at = {(.5,1.03)},
		anchor=south,
		/tikz/every even column/.append style={column sep=0.25cm}
	},
	]
	
	\pgfplotstableread[col sep=comma]{verification_7users.dat}\tbl
	
	\addplot+[smooth] table[y=opt] {\tbl};
	\addlegendentry{Optimal};
	
	\addplot+[smooth] table[y=ANN] {\tbl};
	\addlegendentry{ANN};
	
	\addplot+[smooth] table[y=SCA] {\tbl};
	\addlegendentry{SCA};
	
	\addplot+[smooth] table[y=SCAos] {\tbl};
	\addlegendentry{SCAos};
	
	\addplot+[smooth] table[y=max] {\tbl};
	\addlegendentry{Max.~Power};
	
	\addplot+[smooth] table[y=best] {\tbl};
	\addlegendentry{Best only};
	
	\end{axis}
	\end{tikzpicture}
	\vspace{-2ex}
	\caption{Performance on the test set compared to the global optimal solution, first-order optimal solutions, and fixed power allocations for the scenario with seven users.}
	\label{fig:testset-7users}
	\vspace{-2ex}
\end{figure}
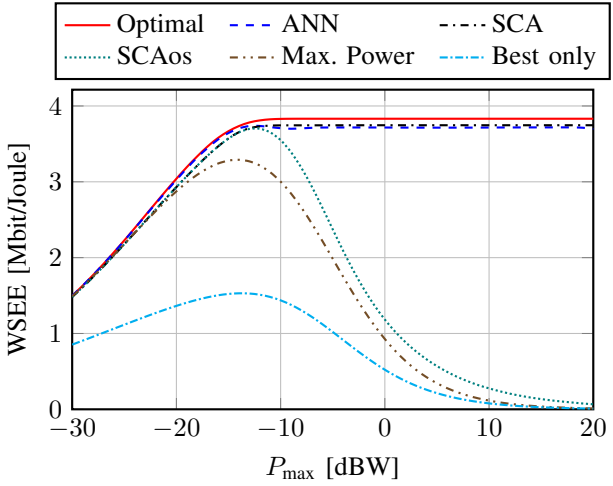

\Cref{fig:testset-7users} shows the performance on the test set compared to the globally optimal solution, first-order optimal solutions and fixed power allocations. Remarkably, the curves look similar to the ones of the smaller problem presented in \cref{fig:testset}. In particular, it can be seen that the \gls{ann} performs virtually optimal in the low \gls{snr} region, outperforming all of the other reference algorithms. At a \cgls{snr} of approximately \unit[-15]{dBW} the \cgls{ann} solution is slightly worse than the globally optimal solution and the \cgls{ann} performs similar to SCA with the two-stage initialization rule discussed in \cref{Sec:Testing}. The results in the previous subsections indicate that this gap can be closed by extended hyperparameter tuning and possibly increasing the size of the training set. 
Thus, it is possible to increase the size of the considered problem and still use the presented framework to obtain a near-optimal solution.

\tikzsetnextfilename{WSEE-7}
\tikzpicturedependsonfile{verification_7users.dat}

\section{Conclusions}\label{Sec:Conclusions}
This work has developed a power control framework for energy-efficient power control in wireless networks. The proposed method is based on a novel \acrlong{bb} procedure wherein specific bounds for energy-efficient problems are derived, which leads to a much faster convergence than other available global optimization methods. Moreover, this complexity reduction allows to train an \gls{ann} using a large dataset of optimal power allocations, which provides a practical power control algorithm, with affordable online complexity. Numerical results have shown that the proposed \gls{ann}-based method achieves near-optimal performance, also being robust against mismatches between the training set and the real testing conditions. Moreover, the proposed \gls{ann}-based method has a much lower complexity than first-order optimal methods, which tackle power control in interference network by solving a sequence of pseudo-convex relaxations, while at the same time yielding comparable or even better performance. 

\appendices

\section{Proof of Theorem~\ref{thm:spca}} \label{sec:proofSpca}
Problem~\cref{opt} has a closed convex feasible set and its objective is a proper, continuously differentiable function. Thus, the result follows from \cite[Thm.~1]{Yang2017} if we can show that \eqref{Eq:EEapprox} satisfies all five technical conditions stated in \cref{Sec:SEQ}.

Conditions~\ref{a1}) and~\ref{a2}) are clearly satisfied, and the boundedness of the feasible set $[\vec 0, \vec P]$ is sufficient for conditions~\ref{a4}) and~\ref{a5}) to hold \cite{Yang2017}.

Condition~\ref{a3}) is equivalent to
$\frac{\partial}{\partial p_k} \widetilde{\text{EE}}_{i}(\vec p; \vec p^{(t)}) \big|_{\vec p = \vec p^{(t)}} = \frac{\partial}{\partial p_k} \text{EE}_{i}(\vec p)\big|_{\vec p = \vec p^{(t)}}$
for all $k = 1, \dots, K$. The \cgls{lhs} can be expressed as
\ifhbonecolumn
	\begin{align}
		\frac{\partial}{\partial p_i} \widetilde{\text{EE}}_{i}(\vec p; \vec p^{(t)})\big|_{\vec p = \vec p^{(t)}}
		&= \Bigg[ w_i \frac{\partial}{\partial p_i} \frac{R_i(p_i, \vec p^{(t)}_{-i})}{\phi_i q_i + P_{c,i}}
				- w_i \frac{\phi_i R_i(\vec p^{(t)})}{(\phi_i q_i + P_{c,i})^2} 
				+ \sum_{j\neq i} w_j
					\frac{\frac{\partial}{\partial p_i} R_j(\vec p^{(t)})}{\phi_j q_j + P_{c,j}} \Bigg]_{\vec p = \vec p^{(t)}} \notag\\
		&= w_i \left( \frac{\frac{\partial}{\partial p_i} R_i(\vec p^{(t)})}{\phi_i q_i + P_{c,i}} - \frac{\phi_i R_i(\vec p^{(t)})}{(\phi_i q_i + P_{c,i})^2} \right)
				+ \sum_{j\neq i} w_j \frac{\frac{\partial}{\partial p_i} R_j(\vec p^{(t)})}{\phi_j q_j + P_{c,j}}.
			\label{eq:diffft}
	\end{align}

\else
	{\small
	\begin{align}
		&\frac{\partial}{\partial p_i} \widetilde{\text{EE}}_{i}(\vec p; \vec p^{(t)})\big|_{\vec p = \vec p^{(t)}}\notag\\
		=& \Bigg[ \frac{\partial}{\partial p_i} \frac{w_i R_i(p_i, \vec p^{(t)}_{-i})}{\phi_i q_i + P_{c,i}}
				-\! \frac{w_i \phi_i R_i(\vec p^{(t)})}{(\phi_i q_i + P_{c,i})^2}
				+\! \sum_{j\neq i}
					\frac{w_j \frac{\partial}{\partial p_i} R_j(\vec p^{(t)})}{\phi_j q_j + P_{c,j}} \Bigg]_{\vec p = \vec p^{(t)}}
			\notag\\
		=&  w_i\! \left( \frac{\frac{\partial}{\partial p_i} R_i(\vec p^{(t)})}{\phi_i q_i + P_{c,i}} - \frac{\phi_i R_i(\vec p^{(t)})}{(\phi_i q_i + P_{c,i})^2} \right)\!
		+ \sum_{j\neq i} \frac{w_j \frac{\partial}{\partial p_i} R_j(\vec p^{(t)})}{\phi_j q_j + P_{c,j}}.
			\label{eq:diffft}
			\raisetag{2em}
	\end{align}}
\fi
The \cgls{rhs} can be expressed as 
\ifhbonecolumn
	\begin{align}
		\frac{\partial}{\partial p_i} \text{EE}_{i}(\vec p)
		&= w_i \frac{\partial}{\partial p_i} \left( \frac{R_i(\vec p)}{\phi_i p_i + P_{c,i}} \right) + \sum_{j\neq i} w_j \frac{\frac{\partial}{\partial p_i} R_j(\vec p)}{\phi_j p_j + P_{c,j}} \notag\\
		&= w_i \left( \frac{\frac{\partial}{\partial p_i} R_i(\vec p)}{\phi_i p_i + P_{c,i}} - \frac{\phi_i R_i(\vec p)}{(\phi_i p_i + P_{c,i})^2} \right) + \sum_{j\neq i} w_j \frac{\frac{\partial}{\partial p_i} R_j(\vec p)}{\phi_j p_j + P_{c,j}}.
			\label{eq:difff}
	\end{align}
\else
	{\small\begin{align}
		&\frac{\partial}{\partial p_i} \text{EE}_{i}(\vec p)
		= w_i \frac{\partial}{\partial p_i} \left( \frac{R_i(\vec p)}{\phi_i p_i + P_{c,i}} \right) + \sum_{j\neq i} \frac{w_j \frac{\partial}{\partial p_i} R_j(\vec p)}{\phi_j p_j + P_{c,j}} \notag\\
		&= w_i \!\left( \frac{\frac{\partial}{\partial p_i} R_i(\vec p)}{\phi_i p_i + P_{c,i}} - \frac{\phi_i R_i(\vec p)}{(\phi_i p_i + P_{c,i})^2} \right)
				+\! \sum_{j\neq i} \frac{w_j \frac{\partial}{\partial p_i} R_j(\vec p)}{\phi_j p_j + P_{c,j}}.
			\label{eq:difff}
	\end{align}}
\fi
When evaluated at $\vec p = \vec p^t$, the \cgls{lhs} and \cgls{rhs} are equal, and thus Condition~\ref{a3}) is satisfied.

\balance
\bibliography{IEEEabrv,IEEEtrancfg,paper.bib}
\end{document}